# On the Road from Edge Computing to the Edge Mesh

Panagiotis Oikonomou[1], Anna Karanika[1], Christos Anagnostopoulos[2], Kostas Kolomvatsos[1]

[1] Department of Informatics and Telecommunications, University of Thessaly, Papasiopoulou 2-4, 35131, Lamia Greece

emails: {paikonom, ankaranika, kostasks}@uth.gr

[2] School of Computing Science, University of Glasgow, Lilybank Gardens 17, G12 8RZ, Glasgow UK

Email: christos.anagnostopoulos@glasgow.ac.uk

**Abstract**

Nowadays, we are witnessing the advent of the Internet of Things (EC) and the involvement of numerous devices performing interactions between them or with their environment and end users. The huge number of connected devices leads to huge volumes of collected data that demand the appropriate processing. The 'legacy' approach is to rely on Cloud where increased computational resources can be adopted to realize the envisioned processing. However, even if the communication with the Cloud back end lasts for some seconds there are cases where problems in the network or the need for supporting real time applications require a reduced latency in the provision of responses/outcomes. Edge Computing (EC) comes into the scene as the 'solver' of the latency problem (and not only). Any processing can be performed close to data sources, i.e., at EC nodes having direct connection with IoT devices. Hence, an ecosystem of processing nodes can be present at the edge of the network giving the opportunity to apply novel services upon the collected data. Various challenges should be met before we talk about a fully automated ecosystem where EC nodes can cooperate or understand the status of them and the environment to be capable of efficiently serving end users or applications. In this paper, we perform a survey of the relevant research activities targeting to support the vision of Edge Mesh (EM), i.e., a 'cover' of intelligence upon the EC infrastructure. We present all the parts of the EC/EM framework starting from the necessary hardware and discussing research outcomes in every aspect of EC nodes functioning. We present technologies and theories adopted for data, tasks and resource management while discussing how (deep) machine learning and optimization techniques are adopted to solve various problems. Our aim is to provide a starting point for novel research to conclude efficient services/applications opening up the path to realize the future EC form.

## 1. Introduction

Nowadays, we are witnessing the huge explosion of the **Internet of Things (IoT)** that incorporates numerous devices interconnected into the same infrastructure [62]. This vast infrastructure gives the opportunity to build/support novel applications in close distance with end users. It is estimated that the potential impact of IoT will be close to $11.1 trillion by 2025 [64] exhibiting its value for all the stakeholders active in various application domains. IoT devices are in the position of interacting with end users and their environment to collect data and perform simple processing activities. IoT has evolved into a network of devices of all types and sizes, e.g., vehicles, smart phones, home appliances, toys, cameras, medical instruments and industrial systems [77]. Data or the outcome of any lightweight processing can be transferred through the adoption of wireless communications to other 'peer' devices or to the Cloud infrastructure where more advanced processing can be the case. Such transfer of data is regulated by the appropriate models and protocols for sharing information and achieving the necessary detection, positioning and control.

Researchers have focused on terms like 'smart' or 'intelligent' associated with IoT, however, it is not clearly defined what is intelligence in the context of IoT and who provides it. Such questions motivated the research community to study the data-centric IoT, data mining in IoT, the interaction of **Artificial Intelligent (AI)** with IoT etc. The majority of the present IoT systems build on a centralized entity (e.g., a server) for computational purposes. This entity is usually placed at the Cloud infrastructure. Low-level IoT devices are used only for sensing purposes and the collection of data while the final decision-making is performed by the central entity. The forthcoming emergence of the **Internet over Everything (IoE)** [67] will extend the capabilities of the 'legacy' IoT, the number of devices and the volumes of data. This explosion will be driven by the evolution of 5G technology, advances in the Cloud infrastructure, the extended use of social media, advances in mobile computing and new trends of data science. It becomes obvious that such an evolution will impose new requirements related to the storage of the huge volumes of data as well as the management of the numerous devices. Eventually, the aforementioned centralized model will be affected by the bottleneck in the processing of the collected data that will increase the latency in the provision of responses as well as by the need for increased bandwidth. Additionally, Cloud may face accessibility challenges, e.g., unstable connections between Cloud and IoT devices. In any case, these problems negatively affect the performance of real-time applications where latency is critical.

It becomes obvious that IoT devices are, usually, lightweight nodes with constrained resources, thus, no advanced processing can take place on them. Consequently, data and the 'light' local knowledge should be transferred to Cloud where increased computational resources are present. However, the transfer of huge volumes of data in the network can have negative effects on the scalability aspect and the required bandwidth [29]. **Edge Computing (EC)** comes into scene to limit the amount of data transported back to the Cloud [74]. EC provides an intermediate infrastructure upon the IoT and below the Cloud that can support services towards reducing the latency in the provision of responses to end users. If we build novel services at the EC, we can limit the connectivity cost and add a layer for storage and processing. Currently, the majority of the collected data in the IoT infrastructure are not used even they could contribute in the production of new knowledge. Considering EC to rely at the middle between the IoT and the Cloud creates multiple challenges. The first challenge deals with interoperability issues, i.e., we need a novel approach to 'aggregate' data coming from various devices. Additionally, we need advanced methodologies and algorithms for the management of data, EC nodes, IoT devices and so on and so forth.

Currently, we are at the early stages of far-reaching and consequential EC revolution to prepare the aforementioned infrastructure for the new, modern, **Edge Mesh (EM)**. EM provides a 'virtual' layer (a computational/processing overlay) that enables the cooperation between ENs of different types to conclude a cooperative infrastructure close to end users [85]. Operators can/should/will open the EC to third-parties, allowing them to rapidly deploy innovative applications and content towards mobile subscribers, enterprisers, and other vertical segments [27]. As things stand today, 'there is no edge' and 'if there is, everything is centralized', Karri Kuoppamaki, VP of T-Mobile US said[1]. In the near future, with the advent of 5G networks, we will be surrounded by a high number of network 'hubs' moving from 4G macrocells to 5G microcells. The future networks may be as ubiquitous as electricity networks while their capacity is exploding to 'host' and process all the data reported by IoT devices, thus, giving the opportunity for 'tiered' data management and limited latency.

In the new era of EC/EM, multiple research questions should be answered like the following: How to define the network and computing model? How to manage/distribute data processing? How to jointly optimize computation? etc. Currently, the distributed processing can be realized in the available EC nodes, however, these nodes are heterogeneous, resource-constraint (compared to the Cloud), and have direct communication with a high number of IoT devices. Other open research issues like the reliability, fault tolerance, **Quality of Service (QoS)** management, security and privacy etc should also be considered. **(Deep) Machine Learning ((D)ML)** and optimization techniques can assist in many aspects for the provision of the appropriate services that will give a boost to the performance of applications with positive impact in end users activities. (D)ML can setup the basis for covering multiple axes of the EC/EM functioning. Nodes placed at the EC/EM can have a 'logical' connection over the physical infrastructure adopting distributed intelligence to conclude a fully automated framework close to IoT devices, thus, to end users. In the intelligent EM, we need a type of 'cooperative computing' for handling the unbalanced computation distribution and lead to better usage of resources, reduced latency and better services as nodes cooperate with each other.

---

[1] https://www.sdxcentral.com/articles/news/operators-strike-realistic-edge-computing-balance/2019/09/

This survey targets to expose the current efforts in creating the new form of the EC, i.e., the EM. Our focus is on the creation of the necessary intelligence to realize an ecosystem of autonomous devices in the EC. The axes of this study are as follows: **(i)** the necessary hardware to support the envisioned processing; **(ii)** the (D)ML and optimization models adopted for realizing the EM; **(iii)** models for data management at the EC; **(iv)** tasks management at the EM; **(v)** autonomous nodes management. We depart from other surveys and focus only on the aforementioned axes. We do not consider 'low' level information putting our efforts in the discussion of the challenges, open issues and the algorithms/models proposed to process data, tasks and nodes at the EC/EM. Through this approach, we pay attention on the support of an intelligent EC/EM ecosystem where numerous nodes can communicate each other, with the Cloud and with the IoT infrastructure to deliver novel applications. We devote a separate section for each axis and provide a description of the relevant efforts in each sub-field. Our aim is to expose the progress related to the management of the aforementioned 'logical' layer that covers the EC physical infrastructure. Finally, we conclude our paper by giving some future directions.

## 2. Edge Computing and Internet of Things

### 2.1 A Layered Architecture

Fig. 1 (retrieved by [14]) shows the generic architecture of the layered approach moving from the IoT infrastructure to the Cloud datacenters. At every layer, a high number of devices can be present capable of supporting the envisioned processing of the collected data and offering a wide set of services. EC refers to the data processing that happens close to where data are collected/produced, i.e. 'at the edge' of the IoT network. In Cloud, processing is realized at the central datacenter with the response returning back to end users after its conclusion. This processing will not take too long (e.g., a couple of seconds), however, in some cases, the provision of the final response could be jeopardized due to, e.g., a network glitch, weak network communications, a high distance between the receiver of the response and the datacenter [80]. EC nodes are located in a close distance with end users, thus, the latency in the provision of responses is limited. As a result, IoT devices are no longer dependent on the connection with the network resulting an autonomous infrastructure.

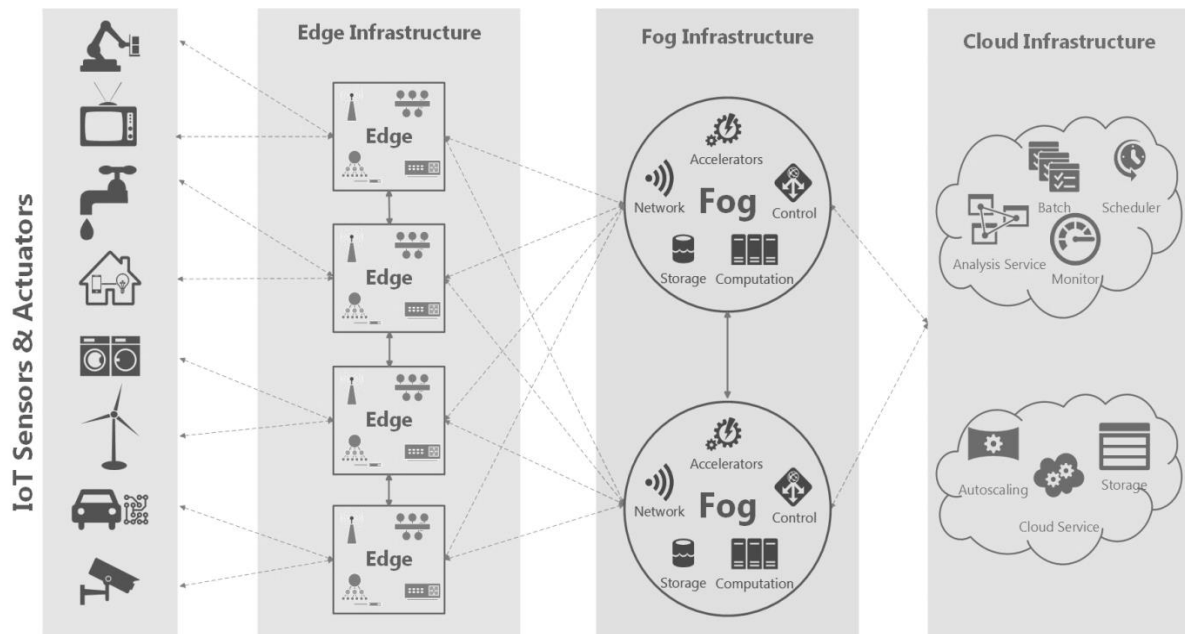

**Fig. 1 Edge-Fog-Cloud Architecture**

Both EC and IoT are a perfect match exposing their complementarity related to the collection, transfer and processing of data. Many times EC is confused with **Fog Computing (FC)** [9]. However, FC exhibits the following differences when compared with EC [3], [69]:

- FC is device independent and aware of the entire fog domain while EC is aware only for every device and a few services;
- FC controls all devices in the domain while EC exhibits limited control;
- FC extends Cloud as a continuum while EC is Cloud unaware;
- FC supports for multiple IoT verticals while EC exhibits no IoT vertical awareness;
- FC nodes are versatile and capable of performing a variety of functionalities while EC nodes are focused on device command and control;
- FC supports end-to-end security while in the EC, the security scope is limited to devices;
- FC supports analytics from multiple devices while EC analytics are oriented to individual devices.

Both, EC and FC, try to keep the processing of data very close to the IoT infrastructure to speed up the provision of responses. The speed in data processing and the immediate provision of analytics are essential in many application domains but they are also the key for transforming industrial processes in many ways. The final target to automate the industrial processing adopting control software, actuators and intelligent decision making.

Moreover, EC provides a better alternative for the efficient management of computational resources. This leaves the room for Cloud administrators to also efficiently manage their resources as a set of functionalities are decoupled. The control of the resources is shifted from Cloud to the edge of the network creating a new layer of administration. Hence, we place a number of management activities close to the source of data giving the opportunity to be aligned with the real needs. A specific example is Edge Analytics [20] adopted by vendors like Intel[2] or Cisco[3] allowing the interested stakeholders to perform a set of pre-processing activities just after the reception of data. The significant aspect is that such an approach can be 'distributed' in the entire ecosystem of EC nodes opening up the path for high quality and novel services. EC nodes can be considered as points where distributed datasets are formulated hosting any desired, however feasible, functionality.

---

[2] https://www.intel.com/content/www/us/en/edge-computing/overview.html
[3] https://blogs.cisco.com/tag/edge-analytics

Consequently, we are enjoying more storage, additional processing points, a fault tolerant approach and enhanced analytics capabilities. It becomes obvious that the processing 'pressure' is taken away from the Cloud infrastructure, however, we have to secure the role of each framework. Another aspect is the opportunity to replicate data and create a fully fault tolerant infrastructure. The following list reports on the advantages of replicating data in multiple distributed datasets:

- **Efficiency**: It is not necessary to always access the 'master' data at the Cloud with positive impact on the performance of applications.
- **Low latency**: If we have direct access on the local data, we can reduce the latency for accessing them in a distant location.
- **Fault tolerance**: By distributing the data across multiple locations, we can easily support fault tolerance being able to recover from any disaster. However, this approach requires the efficient management of replicas and the adoption of mechanisms that manage possible redundancies.
- **Scalability**: Complex processing can be distributed in multiple nodes to benefit from the 'collective' power of the infrastructure. Recall that the available nodes can be provided by different vendors, thus, a powerful model for consistency and heterogeneity management as well as resilience is necessary.

The EM comes to overcome the disadvantages of the EC, i.e.., to facilitate the management of resources and tasks applying the necessary intelligence. It targets to a cooperative model where EC nodes can exchange data, tasks or knowledge in order to perform the desired processing. Eventually, EM acts as an overlay virtual network over EC nodes trying to overcome the problem of constrained resources through a collective intelligence approach. EM is proposed to deal with a computing paradigm that uses a mesh network of EC nodes to enable distributed decision-making, exchange of data and computation among the available nodes. This differs from the existing EC paradigm which usually considers EC nodes as 'simple' nodes responsible to collect and transmit data to the Cloud infrastructure. EM 'imposes' the idea of using EC nodes to enable distributed intelligence in IoT [82], i.e., the cooperation between autonomous entities, intermediate communication infrastructures (local networks, access networks, global networks) and/or Cloud systems to optimally support IoT communication and applications.

Under this rationale, EC/EM and IoT can have a close cooperation to host business processes and support a strong computing paradigm for the future. If we apply intelligence in both, the EC/EM and IoT, we can easily go beyond the state of the art and support the autonomous behaviour of numerous nodes. The envisioned interconnection can be holistically realized either in the horizontal axis or in the vertical one. The vertical approach will refer in the communication between the IoT, the EC/EM, the FC and the Cloud. Hence, we will be able to integrate and distribute the processing activities and power in any direction at will. Eventually, this will lead to the optimization of the use of resources and the ability to dynamically react to end users dynamic requirements. For instance, we can easily control the data flow and the processing needs from one location to another. In any case, such an approach will open up the road for having third parties involved in the provision of advanced services in the discussed ecosystem. In that case, businesses will find more opportunities to expand their portfolios or invest on new services that will increase their revenues and the quality of their products.

Currently, there is a number of companies providing EC services. For instance, Foghorn[4], Swim[5], Juniper[6], Crosser[7], Mutable[8], AlefEdge[9], CLearBlade[10], SSAS[11], Eurotech[12], Edge Intelligence[13] and many others support ML algorithms and analytics for EC/EM. Additional companies like Zededa[14], Edgeworx[15], Affirmed Networks[16], Ori[17], Packet[18], EdgeConneX[19] offer virtualization and orchestration services for EC. Of course the lists are not exhaustive but they just present indicative examples. The aim of all these companies is to have their services interacting with IoT and Cloud while applying specific algorithms and models upon the collected data. They become the first point of expansion for companies working at the provision of IoT platforms. Example companies are: Software[20], QIO[21], Altizon[22], IBM[23], Litmus Automation[24], Exosite[25], Oracle[26], Atos[27]. Dell is the leader of the EdgeX open source project for edge computing[28]. Eurotech offers Everywhere Software Framework[29] for building edge computing applications. ADLINK develops Vortex Edge[30] and Vortex DDS[31] to facilitate the deployment of software on edge gateways.

---

[4] https://www.foghorn.io/
[5] https://www.swim.ai/
[6] https://www.juniper.net/us/en/
[7] https://crosser.io/
[8] http://www.mutable.io
[9] http://www.alefedge.com
[10] http://www.clearblade.com

The aforementioned solutions target to facilitate the incorporation of IoT devices and the collected data to the infrastructure present at higher layers. Hence, these solutions try to offer to customers an integrated approach that includes device to edge, then, to Cloud. The deployment of any software or hardware will be facilitated due to the automatic detection of the requirements and the inclusion of advanced software to realize it. Significant **Key Performance Indicators (KPIs)** deal with the time required to perform any action (e.g., the provision of analytics, the conclusion of the incorporation of any devices) as well as the quality of the final outcome. The challenge here deals with the heterogeneity of products that have to be integrated and communicate in short time. For this, the appropriate APIs and protocols should be implemented together with the necessary wrappers. Already present models (i.e., APIs and protocols offered by Cloud providers) should be also adopted to secure the smooth integration of novel solutions. Example solutions are the initiatives of Amazon[32], Microsoft[33] and Google[34]. The aforementioned companies propose edge solutions connected with their respective IoT platform. Additionally, the majority of the IoT platform vendors (some example are already given above) are also proposing their own edge solutions to expand their portfolio. All of them target to expose a holistic, bottom up, approach to, finally, be able to support intelligent analytics upon the collected data. Of special attention is to facilitate the execution of advanced ML algorithms over huge volumes of data.

Apart from commercial products, one can detect an increased number of efforts to provide open source tools. Such tools try to avoid to be directly connected with a specific vendor, i.e., they offer solutions to avoid the vendor lock-in problem. However, this does not mean that the above discussed companies do not provide open source tools (e.g., IBM). Some open source initiatives are as follows (the list is not exhaustive).

---

[11] https://www.sas.com/en_us/home.html
[12] https://www.eurotech.com/en
[13] https://www.edgeintelligence.com/
[14] https://zededa.com/
[15] https://edgeworx.io/
[16] http://www.affirmednetworks.com
[17] http://www.ori.co
[18] http://www.packet.com
[19] http://www.edgeconnex.com
[20] https://www.softwareag.com/corporate/default.html
[21] https://qio.io/
[22] https://altizon.com/
[23] https://www.ibm.com/ie-en
[24] https://www.ibm.com/ie-en
[25] https://exosite.com/
[26] https://www.oracle.com/
[27] https://atos.net/
[28] https://www.edgexfoundry.org/
[29] https://www.eurotech.com/en
[30] https://www.adlinktech.com/en/Edge-IoT-Solutions-and-Technology
[31] https://www.adlinktech.com/Products/IoT_solutions/Vortex_DDS/Vortex_DDS?lang=en
[32] https://aws.amazon.com/iot/
[33] https://azure.microsoft.com/en-us/overview/iot/
[34] https://cloud.google.com/solutions/iot

**LF Edge Community**[35] is an umbrella organization that aims to establish an open, interoperable framework for edge computing independent of hardware, silicon, Cloud, or operating system. The community brings together leaders in the relevant industry and aims to create a common framework for hardware and software standards. Additionally, it aims to expose the best practices that are significant to sustain current and future generations of IoT and edge devices. The community fosters collaboration and innovation across the multiple industries, i.e., industrial manufacturing, cities and government, energy, transportation, retail, home and building automation, automotive, logistics and health care — all of which stand to be transformed by edge computing. A high number of companies active in the EC/EM domain are members of the effort (see https://www.lfedge.org/members/).

**Akraino Edge Stack**[36] is a set of open infrastructures and application blueprints for the EC, spanning a broad variety of use cases, including 5G, AI, Edge IaaS/PaaS, IoT, for both provider and enterprise edge domains. These blueprints are proposed by the Akraino community (part of the LF Edge) and focus exclusively on the edge in all of its different forms. Hence, the community tries to setup the basis for defining blueprints for all the aspects of the edge infrastructure. The connection between the blueprints is secured by the community and the testing procedures to deliver solutions that can be adopted as-is.

**Eclipse ioFog Project**[37] is a complete edge computing platform that provides all of the pieces needed to build and run applications at the edge at enterprise scale. The project provides abstractions to manage the diversity and complexity of edge hardware. Hence, a software 'cover' is adopted to support the necessary functionalities to avoid problems related with the underlying heterogeneity of devices and software. The project also targets to the management and orchestration of edge microservices performed by the dedicated ioFog Controller and its supporting set of components.

The **OSF Edge Computing Group**[38] targets to define infrastructure systems required to support applications distributed over a broad geographic area, with potentially thousands of sites, located as close as possible to discrete data sources, physical elements or end users. All these applications can communicate over wireless communications. Another goal of the group is to detect use cases, develop requirements, and produce viable architecture options for evaluating new and existing solutions, across different industries and global constituencies, to enable development activities for Open Infrastructure and other Open Source community projects to support EC use cases.

**StarlingX**[39] is a complete Cloud infrastructure software stack for the edge used by the most demanding applications in industrial IoT, telecom, video delivery and other ultra-low latency use cases. The approach proposed by the StarlingX is oriented around the provision of a container-based infrastructure for edge implementations in scalable solutions that is ready for production. The focus is on easy deployments, low touch manageability, rapid response to events and fast recovery. The solution is tested and released as a complete stack, thus, it ensures the compatibility among diverse open source components.

**CORD (Central Office Re-architected as a Datacenter)**[40] is a project that intents the transformation of EC into an Agile service delivery platform enabling the operator to deliver the best end-user experience along with innovative next-generation services. The proposed platform builds upon Software Defined Networks (SDNs), Network Functions Virtualization (NFV) and Cloud technologies to build agile datacenters for the network edge. Integrating multiple open source projects, CORD delivers a cloud-native, open, programmable, Agile platform for network operators to create innovative services.

---

[35] https://www.lfedge.org/#
[36] https://www.lfedge.org/projects/akraino/
[37] https://iofog.org/
[38] https://www.openstack.org/edge-computing/
[39] https://www.starlingx.io/
[40] https://www.opennetworking.org/cord/

**EdgeX Foundry**[41] is a vendor-neutral open source project hosted by The Linux Foundation building a common open framework for IoT edge computing. The main focus of the project is the provision of an interoperability framework hosted within a full hardware and OS-agnostic reference software platform to enable an ecosystem of plug-and-play components that unifies the marketplace and accelerates the deployment of IoT solutions.

**KubeEdge**[42] is an open source system for extending native containerized application orchestration capabilities to hosts at Edge. It adopts Kubernetes[43] (an open-source system for automating deployment, scaling, and management of containerized applications) and provides a fundamental infrastructure support for network, application deployment and metadata synchronization between Cloud and edge. KubeEdge is licensed under Apache 2.0. and free for personal or commercial use.

## 2.2 Intelligence at the Edge of the Network

The development of intelligent applications in IoT has gained significant attention in recent years. Smart devices/sensors can be interconnected each other or with the back end systems to enable various services in multiple domains. Several intelligent technologies have emerged to ensure the proper functioning of IoT devices and their incorporation into the above discussed ecosystem. For instance, the efficient management of software updates of IoT devices aiming at the minimization of the conclusion time and the maximization of their and network's performance is the subject of some recent studies [44], [48], [51]. Cloud provides many benefits to the IoT infrastructure, e.g., high-performance computing, storage resources, processing and analysis of huge volumes of data. Hence, the IoT can be robust, smart and self-configuring. The forthcoming IoE will extend the capabilities of the 'legacy' IoT focusing on the intelligent communication between people, process, data and things [3]. Based on the above, we can easily identify the need for transformation of the interactions between all these 'entities' leading to new types of communications like **Machine to Machine (M2M)** and **Person to Machine (P2M)**. In this new environment, we can detect new requirements for concluding an intelligent ecosystem, e.g., **(i)** advanced models for devices management; **(ii)** intelligent schemes for supporting communications; **(iii)** smart frameworks for the management, processing and storage of humongous volumes of heterogeneous data generated at the network edges.

As stated above, Cloud technologies face some accessibility challenges when providing services to end-users. An example concerns mobile clients who can move among different places, yet require Cloud services with minimum cost and limited response time. It becomes obvious that mobile communications can be heavily affected by problems in the communication channel. This, in a sequential order, can create severe problems in real-time applications where latency is critical. Several EC/EM technologies, originating in different backgrounds, have emerged to decrease latency and support the massive machine type of communication. However, there is an intense need to support the EC/EM nodes with intelligent services to overcome all the aforementioned challenges.

IoT devices will, now, communicate with EC nodes in short distance relying on them to enjoy advanced services and efficiently serve end users. This means that EC nodes should 'convey' the necessary knowledge, processing and decision making mechanisms to serve IoT devices and the Cloud infrastructure. In any case, EC/EM technologies face various challenges and open research issues. The following paragraphs report on a set of challenges that are critical for the transformation and the delivery of the new EC/EM.

---

[41] https://www.edgexfoundry.org/
[42] https://kubeedge.io/en/
[43] https://kubernetes.io/

The future intelligent EC/EM (coined here as the 'new edge') will involve numerous autonomous entities capable of understanding their status, the status of the environment and their peers taking actions on the fly to efficiently serve the desired applications. This ecosystem targets to secure the efficient execution of any application learning how to: behave, collaborate, exchange data and tasks, process data and tasks, forecast abnormal situations, select the best strategy to work, react in potential errors or demands in an autonomous manner. It is the appropriate time to provide AI-enhanced autonomous entities at the edge of the network that future applications demand. We have to combine and adapt already present solutions, currently lying isolated, together with new models, algorithms, methods and technologies to holistically target to autonomous entities capable of efficiently serving end users & applications. Apart from the latency minimization (a pivotal aspect in provision of real time services), we should also deal with the QoS and, more importantly, the **Quality of Experience (QoE)**. Future applications will take seriously into consideration the latency combined with the efficiency and the quality of the delivered services. It is not enough to deal only with the latency but QoS and QoE should be also taken into consideration at the same time. Various KPIs will be adopted to depict QoS and QoE, thus, intelligent monitoring modules should be created to decide upon these KPIs. Decisions will be related to the configuration, the cooperation and, in general, the management of EC/EM nodes. KPIs will cover all the aspects of EC/EM nodes functioning being related to the performance of the network, users' satisfaction and potential overheads associated with deployment/migration/replication of the required processing tasks and data. An efficient edge infrastructure is the key challenge for the envisioned IoT applications; it is critical with the growing demand for energy-hungry applications, such as video streaming, Augmented Reality (AR) and 3D gaming.

## 3. Hardware Requirements

EC/EM functionalities are provided upon the relevant hardware to perform the desired processing or communication. A survey on 'low' level characteristics of the necessary hardware is presented in [15]. In this section, we perform a 'high' level review with limited technicalities to setup the basis for understanding what we need to efficiently support EC/EM and deploy solutions from the hardware perspective.

The difference of the EC/EM (compared to Cloud) is that it is distributed across a high number of devices in different locations. This adds a burden on the connection with all the appropriate devices together with their configuration and management. Additional requirements deal with the data present at the distributed locations, e.g., there should be an efficient storage mechanism across different devices, advanced replication and consistency checking models as well as efficient strategies for baking up the available data. In any case, a strong scheme for implementing the coordination with the Cloud back end for data management could increase the performance and assist in solving the aforementioned problems. However, the EC/EM hardware infrastructure should be efficiently connected with the Cloud 'premises' to facilitate the easy communication between the two frameworks when executing the proposed algorithms/models.

The first key component of EC/EM is the communication support. EC/EM should support the connection with the IoT devices (e.g., sensors, actuators) based on already proposed technologies and protocols. The most common protocols for communication are as follows: Bluetooth[44], ONVIF[45], Z-Wave[46], ZigBee[47], LoRa[48], KNX[49], Siemens S7[50], HomeConnect[51], Modbus[52], EnOcean[53], BACnet[54],

---
[44] https://www.bluetooth.com/
[45] https://www.onvif.org/
[46] https://www.z-wave.com/
[47] https://zigbeealliance.org/
[48] https://lora-alliance.org/
[49] https://www.knx.org/knx-en/for-professionals/index.php
[50] https://wiki.wireshark.org/S7comm

OPC[55]. Apart from that, EC/EM nodes should support data and device management protocols like MQTT[56], CoAP[57], AMQP[58], Websockets[59], TR-069[60], OMA-DM[61]. This way, IoT devices will be facilitated to transfer data in an upwards mode and cooperate in processing activities.

The processing at the EC/EM can be performed by dedicated hardware in various forms, e.g., gateways, routers, microservers, etc (see Fig. 2). Currently, various companies propose the use of a set of high quality hardware with small size, however, with increased computational resources. The aim is to efficiently support advanced processing at the edge of the network upon the large scale data streams. The first example deals with the Graphical Processing Unit (GPU) [76]. A GPU is a chip dedicated to perform advanced calculations very quickly. Usually, GPUs are adopted for rendering images alleviating the main processor from these activities. For instance, NVIDIA offers multiple solutions for incorporating GPUs in other devices[62].

**Field Programmable Gate Arrays (FPGAs)**[63] are semiconductors IC where the majority of the electrical functionality in the device can be altered by the design engineer, during the PCB assembly process, or after the adoption of the equipment. FPGAs offer many advantages like reducing the latency in the computation, they can be directly connected to inputs and offer a high bandwidth. FPGAs exhibit worse compatibility than the GPUs, however, they require limited programming skills.

**Coral Edge TPUs**[64] are Google's proposal for edge computing. More specifically, the Dev Board is a single-board computer that is ideal for performing fast (D)ML inferencing. TPUs/Dev Boards can be adopted to prototype embedded systems and scale them to production using the on-board Coral System-on-Module (SoM) combined with a custom PCB hardware. The Edge TPU coprocessor is capable of performing 4 trillion operations (tera-operations) per second (TOPS), using 0.5 watts for each TOPS (2 TOPS per watt).

The **EGX Edge Computing Platform**[65] is proposed by NVIDIA to deliver accelerated AI computing to the edge with an easy-to-deploy Cloud native software stack, a range of validated servers and devices, and a vast ecosystem of partners who offer EGX through their products and services. The EGX hardware portfolio starts with the power-efficient NVIDIA Jetson Family, which includes the small but mighty Jetson Nano and Xavier NX providing between 0.5 to 21 trillion operations per second (TOPS) for tasks such as image recognition and sensor fusion. Additionally, the hardware scales to a full rack of NVIDIA T4 servers, delivering more than 10,000 TOPS to serve hundreds of users with real-time speech recognition and other complex AI experiences.

The **Raspberry Pi**[66] is a series of small single-board computers to, initially, promote teaching of basic Computer Science. It is widely used in applications developed for various research domains because of its low cost and portability. Several implementations of Raspberry Pis have been released so far. Pis feature a Broadcom System on a chip (SoC) with an integrated ARM-compatible CPU and on-chip GPU. CPU speed ranges from 700 MHz to 1.4 GHz for the Pi 3 Model B+ or 1.5 GHz for the Pi 4; on-board memory ranges from 256 MB to 1 GB random-access memory (RAM), with up to 8 GB available on the Pi 4. Secure Digital (SD) cards in MicroSDHC form factor (SDHC on early models) are used to store the operating system and program memory.

---

[51] https://www.home-connect.com/global
[52] https://modbus.org/
[53] https://www.enocean.com/en/
[54] http://www.bacnet.org/
[55] https://opcfoundation.org/
[56] http://mqtt.org/
[57] https://coap.technology/
[58] https://www.amqp.org/
[59] https://tools.ietf.org/html/rfc6455
[60] https://www.broadband-forum.org/download/TR-069_Amendment-6.pdf
[61] https://www.openmobilealliance.org/wp/Overviews/dm_overview.html
[62] https://www.nvidia.com/en-gb/graphics-cards/
[63] https://www.xilinx.com/products/silicon-devices/fpga/what-is-an-fpga.html
[64] https://coral.ai/
[65] https://www.nvidia.com/en-us/data-center/products/egx-edge-computing/
[66] https://www.raspberrypi.org/

Micro servers can be also adopted for hosting the collected data and performing the desired processing for delivering analytics. According to [68], the micro server market was valued at USD 39.71 billion in 2019 and is expected to reach USD 67 billion by 2025, at a CAGR of 9.11% over the forecast period 2020 - 2025. This growth will be driven by the expansion of EC/EM and the applications it hosts. The market will be enhanced by the adoption of M2M learning and IoT-enabled devices that create the need for more advanced services. Micro servers will be able to host services related to the management of huge volumes of structured and unstructured data or trivial workloads. The presence of a pre-installed operating system in micro servers will facilitate the deployment of new services and expose the infrastructure to small and medium size enterprises. However, the lack of standardization requires more intensive efforts towards the easiness of the integration of multiple heterogeneous models and algorithms.

**Communication support**

| Transmitter | Concentrator | Gateway | Wireless Term. Unit |

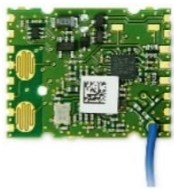 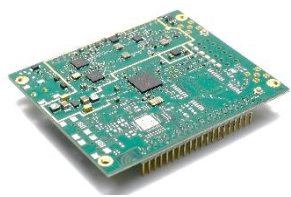 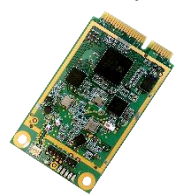 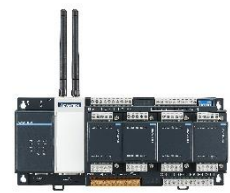

**FPGA / GPU**

EDGE Spartan 6 FPGA[67]   Intel Aria 10 FPGA[68]   NVIDIA EGX™ A100 GPU[69]

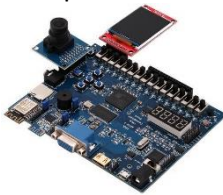 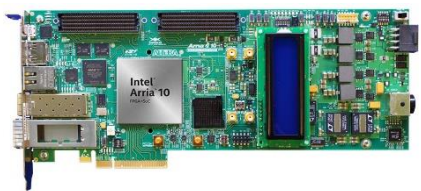 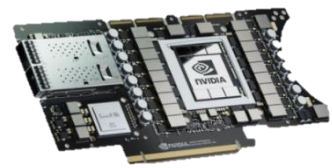

**AI & Machine Learning solutions / Dev boards**

NVIDIA® Jetson™ TX2[70]   TPU Dev Board[71]   Raspberry Pi[72]

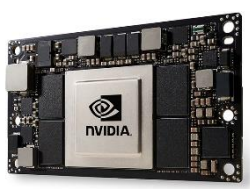 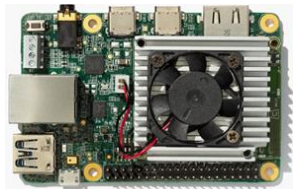 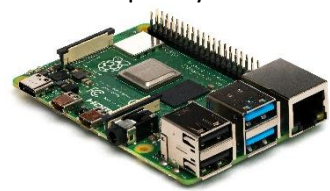

**Servers**

Intel Fog Reference Design unit[73]   Azure IoT Edge[74]   Lenovo Edge Server[75]

---

[67] https://allaboutfpga.com/product/edge-spartan-6-fpga-development-board/
[68] https://www.intel.com/content/www/us/en/products/programmable/fpga/arria-10.html
[69] https://www.nvidia.com/en-au/data-center/products/egx-converged-accelerator/
[70] https://developer.nvidia.com/embedded/jetson-tx2
[71] https://coral.ai/products/
[72] https://www.raspberrypi.org/
[73] https://www.reflexces.com/wp-content/uploads/2018/11/fog-reference-design-overview-guide.pdf
[74] https://azure.microsoft.com/en-us/services/iot-edge/
[75] https://www.lenovo.com/us/en/data-center/servers/edge/thinksystem-se350/p/77XX6DSSE35

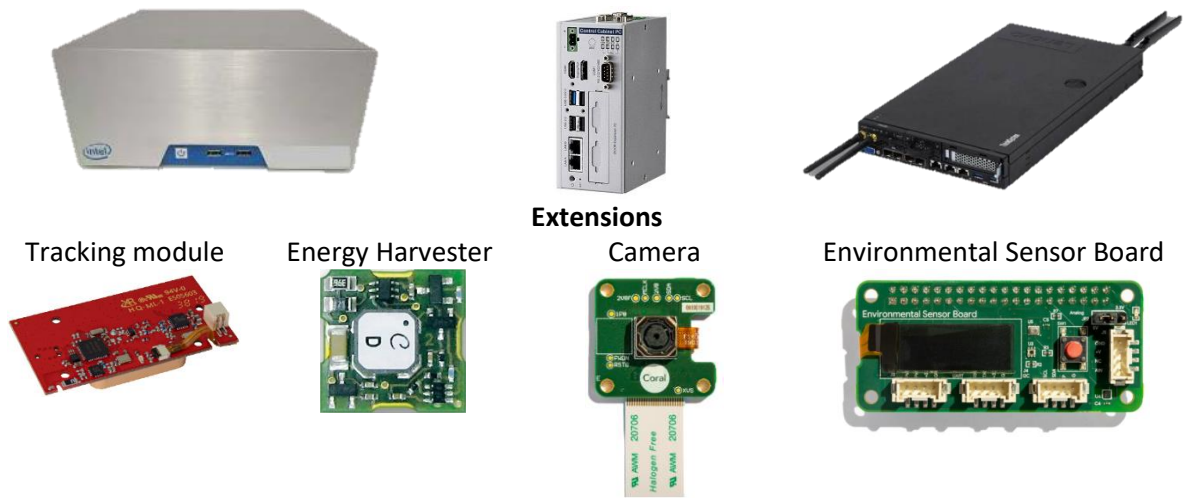

Figure 2. Hardware examples per processing category

## 4. (Deep) Machine Learning Models

**Machine Learning (ML)** refers to the process that a machine has to go through in order to learn a certain behavior and then recognize it, replicate it, or predict new ones without needing to be explicitly programmed on how to do that. ML algorithms receive a considerable amount of data from which they try to extract information. The extraction, also known as training of the algorithm, involves the use of mathematical models or mechanisms that focus on eliminating the errors made by the final model. Some algorithms take into consideration statistics about the dataset to produce the final model, whereas others start with a guess and improve the model incrementally as they process each of the provided dataset's entries. Traditional ML methods are not computationally efficient or scalable enough to handle both the characteristics of big data (e.g., large volumes, high speeds, varying types, low value density, incompleteness) and uncertainty [30].

### 4.1. Machine Learning Models

There are various types of machine learning algorithms, such as supervised, unsupervised, and semi-supervised learning, reinforcement learning, feature learning and association rules algorithms.

**4.1.1 Supervised Learning**

Supervised learning algorithms take as input a training dataset whose every entry consists of a vector of features along with a target value or label. Based on the dataset, a mathematical model is built that works as follows; when receiving as input the feature values of an instance of the dataset, it attempts to produce the corresponding value or label. In essence, a function is created that receives as input the values for specific features and produces a value or decision as its output. The discussed function's error rate has been minimized to be congruous to the training dataset. Supervised learning can be performed with regression and classification algorithms.

A linear regression algorithm models the target value of the training dataset as the linear combination of its features. that predicts an output value based on the input feature values. More specifically, every feature is assigned a coefficient and there is also a bias, such that $y = a_1 x_1 + a_2 x_2 + \cdots + a_n x_n + b \Leftrightarrow y = \boldsymbol{a} \cdot \boldsymbol{x} + b$, where $y$ represents an instance's target value, $\boldsymbol{x}$ is a vector containing the features of the instance, $\boldsymbol{a}$ is the vector containing the coefficients corresponding to the features $\boldsymbol{a}$ and $b$ is the bias. The problem is finding the right coefficients and bias that result in the most accurate predictions of the output value given the input values; this is achieved by minimizing the **Mean Squared Error (MSE)** of the predicted values against the true values of the dataset instances.

Classification, on the other hand, is the process of categorizing an instance of data to the most appropriate of a number of available classes. Classification can be performed after the training of a Support Vector Machine (SVM), a logistic regression classifier, a Naïve Bayes classifier, a decision tree, a k-nearest neighbors algorithm, a neural network, or other algorithms. Support vector machines depict all the instances of a training dataset in a multi-dimensional space and then try to create one or more hyperplanes that divide the instances into the two or more separate classes. A Naïve Bayes classifier depends on the Bayes theorem, i.e. $P(A|B) = \frac{P(B|A)P(A)}{P(B)}$, to predict the class label of an instance. In particular, the application of the Bayes theorem is the following: $P(C_k|x) = \frac{P(x|C_k)P(C_k)}{P(x)}$, meaning that the probability of an instance's class to be $C_k$ given that the feature vector is $x$ is equal to the probability of the feature vector $x$ occurring in class $C_k$ multiplied by the probability of an instance belonging to class $C_k$ and divided by the probability of an instance being described by the feature vector $x$. A decision tree is created from the root all the way down to the leaves by dividing the dataset based on the criteria that divide it most efficiently; all leaves are affiliated with a class and when a prediction needs to be made an instance starting from the root follows the path to the appropriate leaf based on its feature values and node rules. The k-nearest neighbors (k-NN) algorithm predicts an instance's class by taking into consideration its k (specified) closest neighbors in the multi-dimensional space; the instance's distance from all other instances of the training dataset is computed to choose the k nearest ones and come to a decision about its class. A logistic regression classifier works very similarly to the linear regression algorithm, with the only difference being that at the end it classifies the instance to a class based on the computed linear combination of the feature values.

### 4.1.2 Unsupervised Learning

Unsupervised learning, also known as self-organization, is a process which takes as input a dataset that is unlabeled and tries to discover its underlying structure. Unsupervised learning can be achieved through the use of clustering, anomaly detection and other algorithms.

A clustering algorithm receives a set of unlabeled data and attempts to divide them into groups. Maybe the most well-known clustering algorithm, the k-Means clustering algorithm sets a specified number of cluster centroids based on random data entries and then classifies each of the rest of the data entries to the cluster whose centroid is closest to it. After all the entries have been classified, the cluster centroids are recomputed based on their members and all the data are classified once again. This process goes on until the cluster centroids become stable or a maximum number of iterations, also known as epochs, is reached. Hierarchical clustering attempts to form a hierarchy of clusters, and there are two different methods of achieving that: agglomerative or divisive hierarchical clustering. The former strategy constitutes a bottom-up approach; it begins with all of the data entries making up their own cluster and the hierarchy is created by the merge of pairs of clusters, until the top layer is reached where all clusters have been merged into one. The latter strategy is a top-down approach; in the beginning, all data entries make up a single cluster and the hierarchy is created by recursively splitting a cluster until all clusters contain only one entry. The resulting tree is depicted in a dendrogram. Cutting it at a specified height gives one the corresponding clusters. Another popular clustering algorithm is DBSCAN; it divides data points into clusters based on the density of the different groups of points. In particular, groups of data points that are close enough to a specified number of other points, along with any other points in their reach, form clusters; the rest of the points are classified as noise points. OPTICS is an algorithm that combines hierarchical clustering with density-based clustering.

Anomaly detection involves the training of an algorithm to recognize a set of patterns or behaviors as normal and the rest as outliers. Anomalous items are also called novelties, deviations, noise, or exceptions. Density-based techniques, such as k-nearest neighbor, local outlier factor, and isolation forest are very popular in the literature; data points that are not members of high-concentration clusters, are by default anomalous. One-class SVMs are another option, seeing as they form a hyperplane that divides most of the data from the noise. In general, cluster analysis methods can also be used to perform outlier detection. Bayesian networks and Hidden Markov models, as well as, many more algorithms also provide solutions to the anomaly detection problem.

### 4.1.3 Reinforcement Learning

**Reinforcement learning (RL)** involves the existence of an environment and an agent. Unlike supervised learning, RL does not need a labeled dataset to train an agent; it rather attempts to find a balance between exploring the uncharted territory and exploiting the current knowledge. In particular, the environment has a state while the agent repeatedly takes actions in order to reach an objective. Every time the agent takes an action, the environment gives back a reward along with its next state; the reward reflects the agent's progress towards its goal. This happens until the environment reaches a final state; either that the agent has achieved its objective or not. RL utilizes dynamic programming —breaking down a problem into subproblems and recursively solving it— in combination with the **Markov Decision Process (MDP)**, having the agent's actions be dictated by the probability that they will lead to a successful outcome, all the while trying to maximize its cumulative reward. The aforementioned probability is determined by a policy that the agent employs and is dependent on the agent's current state. Furthermore, a value is defined as the long-term expected return of a state given a policy and a Q-value, also known as action-value, is defined as the long-term expected return of a state-action pair given a policy.

The most well-known algorithm to implement RL is Q-learning, which is based on the Bellman equation [1] and gives no guarantee that a solution will be found. Applying the Bellman equation to the agent-environment system, we have: $v(s) = \mathbb{E}[R_{t+1} + \lambda v(S_{t+1})|S_t = s]$ where $v(s)$ is the value of a state $s$, $R_{t+1}$ refers to the immediate reward, $\lambda$ is the discounting factor of future state values and $\mathbb{E}$ depicts the expectation. In the form of Q-value, the equation is transformed to $Q^\pi(s, a) = \mathbb{E}_{s'}[r + \lambda \max_{a'} Q^\pi(s', a') |s, a]$, where $Q^\pi(s, a)$ is the action-value of a state-action pair. The goal is to find the optimal/maximum Q-value; Q-Learning uses a greedy policy to achieve that. Another algorithm is State-Action-Reward-State-Action (SARSA), which is quite similar to Q-learning. Their difference is that SARSA computes the optimal Q-value given the action that is performed based on the current policy instead of the greedy policy. Lastly, the policy gradient method relies upon the optimization of the agent's policy in relation to the long-term cumulative reward by performing gradient ascent, using the REINFORCE algorithm as well as the actor-critic architecture. This approach avoids the lack of guarantee of a value function, as well as, the complexity that the previous approach entails.

### 4.1.4 Algorithms Categorization

In Table 1, we provide a categorization of the above discussed algorithms and present their characteristics. For each category, we also provide representative models together with their implementation complexity and usage.

**Table 1 Machine Learning models**

| ML Models | Supervised Learning | Unsupervised Learning | Reinforcement Learning |
|---|---|---|---|
| **Type of data** | Labeled | Unlabeled | No predefined data |
| **Training** | External supervision | No supervision | No supervision |
| **Complexity** | Simple | High | High |
| **Usage** | Predict outcome/future by mapping labelled data inputs to known outputs | Find hidden structure in data and discovers the output | Learn series of actions using trial-and-error method |

| **Algorithms** | *Regression* <br>• Logistic Regression <br>• Decision Tree Regression <br>• LASSO Regression <br>• Ridge Regression <br>• ElasticNet regression <br><br>*Classification* <br>• support-vector machine (SVM), <br>• logistic regression classifier, <br>• Naïve Bayes classifier, <br>• decision tree, <br>• k-nearest neighbors <br>• Neural network | *Clustering* <br>• K-means clustering <br>• Hierarchical clustering <br>• DBSCAN <br>• OPTICS <br><br>*Anomaly detection* <br>1. K-NN (k nearest neighbors) <br>2. local outlier factor <br>3. isolation forest <br>4. One-class SVMs <br>5. Bayesian networks <br>6. Hidden Markov models | • Q-learning <br>• State-Action-Reward-State-Action <br>• QV <br>• ACLA |
|---|---|---|---|

## 4.2 Deep Learning Models

Another way to approach ML algorithms is to form topologies of nodes, also known as neurons, that collaborate to achieve learning. Neural network algorithms are characterized as deep learning algorithms because they can model especially complex behaviors and patterns, in contrast to other approaches. Their setting enables them to delve deep into the available data and extract comprehensive and accurate knowledge.

### 4.2.1 Supervised Learning

Neural networks provide a more sophisticated approach to linear regression. In fact, neural networks have an input layer, an output layer, and a number of hidden layers; the input layer has as many nodes as the features of the dataset and the output layer has as many nodes as the output values we want to predict. Had there been no hidden layers, a neural network would seem like a simple linear regressor; now, the nodes of each layer receive as input the output of the previous layer and compute their own output which propagates in a forward direction until the outcomes of the output layer are computed. In its training phase, an instance's prediction error is computed at the output layer and is gradually 'back-propagated' until all the layers' weights and biases are optimized, each node optimizing its coefficients quite similarly to how a single linear regressor does. This kind of neural network is called a **multi-layer adaptive linear element (MADALINE)**.

Apart from the basic neural network manner of training, there are also the **radial basis function (RBF)** neural networks. A node in such a network behaves a lot like the k-NN algorithm, in that it receives as input the distance from a set of points in the multi-dimensional space. It essentially functions based on the idea that similar inputs produce similar output values. Another entirely different approach to neural networks is the one presented by **convolutional neural networks (CNNs)**. In their case, the input has 3 dimensions and usually represents the pixels of an image. One type of layer in CNNs consists of filters being convoluted with the output table of the previous layer. Pooling is another type of layer in CNNs; it produces a smaller table that condenses the contents of the previous layer's output table. Furthermore, another layer may contain an activation function being applied to the previous layer's output table. Lastly, a final set of layers is often fully connected, resembling a multi-perceptron layer.

Neural network classifiers also work in a similar fashion to their regression counterparts; just like with the logistic regression algorithm, an activation function decides what class the input instance belongs to, based on the output value of the neural network. An appropriate activation function can also be applied to the output of any one node, before that is used for further computation in the next layer. Some activation functions are rectifier linear unit (ReLU), the hyperbolic tangent, the logistic/sigmoid function, the softplus function and more.

All the previous networks can be either static or dynamic. Static networks depend only on the current input to the network to compute its output. In contrast, dynamic networks depend not only on the current input to the network but also on the network's past inputs, outputs, or states. Dynamic networks contain delay lines that have a forward or backward (recurrent) direction or both. Therefore, the order in which the inputs enter the network can make a huge difference to the output. One of the most well-known recurrent networks is the **long short-term memory (LSTM)**.

### 4.2.2 Unsupervised Learning

Neural network topologies have been designed to perform unsupervised learning. First off, autoencoders are neural networks that receive a set of data and (learn to) encode them without following a specific preprogrammed algorithm. Their main advantage is their ability to perform efficient dimensionality reduction, keeping only the useful information and discarding of the rest. Deep belief networks are trained to reconstruct an input to their topology given an output result; they consist of visible and hidden units where the former represent the training dataset features and the latter an output vector. To train such a network, each of the layers serves as input to the other; the two vectors are computed back and forth until the weights are trained to turn information from one format to the other, quite similarly to autoencoders. Hebbian Learning is based on Hebb's rule which states that when one cell A repeatedly fires another cell B, then A's efficiency is increased [35] or else 'neurons that fire together wire together' [13]. Applying this rule to artificial neural networks, a specified number of clusters can be created by sequentially combining each entry of a dataset with the representative of the cluster that is closest to it, until all cluster representatives reach their final form. A **generative adversarial network (GAN)** can produce a dataset that has the same statistics as its training dataset. It consists of two neural networks that are adversaries to one another; the generative and the adversarial network. The former generates candidate data, while the latter evaluates their plausibility and/or fitness into the distribution of the original data; candidates that are classified as real data are saved to the new dataset. A **self-organizing map (SOM)** is trained to create a low-dimensional (generally two-dimensional) representation of a high-dimensional dataset, essentially performing dimensionality reduction. A SOM's training is carried out with the use of competitive learning - a variant to Hebbian learning - instead of back propagation, and a neighborhood function that helps maintain the input data's properties.

### 4.2.3 Reinforcement Learning

In RL, neural networks are not typically the main attraction. On the contrary, they are used to estimate the optimal Q-value for an agent that would otherwise have to be computed. The deep Q network (DQN) algorithm, for instance, trains a neural network consisting of 2 convolutional and 2 fully connected layers on the target Q-value based on the Q-learning update equation. Subsequently, the DQN's complexity is much lower than that of the original Q-learning algorithm. Moreover, deep deterministic policy gradient (DDPG) is the classic policy gradient algorithm taken a step further to include a neural network, thus becoming faster to train.

### 4.2.4 (D)ML Algorithms Categorization

In Table 2, we provide a categorization of the above discussed algorithms and present their characteristics. For each category, we also provide representative models together with their usage.

**Table 2 Deep Neural Networks characteristics**

| Deep Neural Networks | Activation function | Dataflow | Usage |
|---|---|---|---|
| **Supervised Learning** | | | |
| **MADALINE** | Sign | Feed forward | Classification, Regression |
| **Radial basis functions** | Radial basis | Feed forward | function approximation, time series prediction, classification |
| **Recurrent** | Sigmoid Hyperbolic tangent | Any direction | language modelling, Long short-term memory |
| **Convolutional** | ReLU | Feed forward | computer vision, speech recognition |
| **Unsupervised Learning** | | | |
| **Autoencoders** | Sigmoid function ReLU Softplus | Feed forward | dimensionality reduction, structured prediction, anomaly detection |
| **GAN** | ReLU | Feed forward, Adversarial process | Fashion, art and advertising, Video games |
| **SOM** | None | Feed forward | dimensionality reduction |
| **Deep belief networks** | Hyperbolic tangent | Any direction | Motion-capture, Image/video recognition |
| **Reinforcement Learning** | | | |
| **Deep QN** | ReLU | Feed forward | Q-value approximation |

## 5. Data Management at the Edge

### 5.1 High Level Description

In recent years, data are continuously being collected by sensors and end users, amounting to huge volumes which were traditionally pulled to the Cloud. Today, the edge of the network can assist to alleviate the traffic by playing the role of the mediator. As [81] proposes, the edge can pull the data at first, then, send them to the Cloud at an acceptable rate. The more advanced alternative would be to store data at the edge and have them be closer to users or store them both at the edge and the Cloud for security reasons.

Data management models are essential in an EC/EM environment. EC can only be efficient if data are stored at an appropriate node with respect to their popularity. Accordingly, the data provision latency will usually be low if users requesting them are close to the node that they are stored at. Nevertheless, data have to be distributed in a uniform manner as much as possible. If not, the danger of creating big data centers that start to resemble Cloud infrastructures arises.

The authors of [33] propose a fitting solution to the problem by taking into consideration four important factors; the spatio-temporal locality of range queries, the corresponding information and type of data, the incessant creation and collection of data, as well as the requirements for increased availability of stored data. More specifically, (i) they index data by their spatio-temporal features and place them close to the interested clients, while also (ii) creating replicas on other edge nodes and Cloud, (iii) allowing application developers to define groups of nodes which can balance their loads among each other, and (iv) applying a time-to-live eviction policy, data aggregation and compression to the incoming data. Therefore, this system decreases data latency, is fault-tolerant, strives for load balance and de-escalation of data hotspots, and keeps resource requirements at a minimum, presenting an efficient approach to data management at the edge.

In [84], an architecture for the distributed storage of real-time machine vision data at the edge is proposed. The authors argue that data storage architectures for the edge of the network should be designed keeping in mind the application that uses the specific data. Feature vectors that describe the objects in an image are transferred to the edge at a high rate, whereas images (100x larger in size) whose primary purpose is to be archived and can afford to lose some accuracy are transferred less often, and in case they are similar enough to previous ones they are discarded. The architecture's main idea is to take the advantage of the difference between the data types' latency requirements and the tolerance in images' accuracy loss. Such architecture allows for the efficient management of data related to machine vision at the edge, seeing as it avoids placing useless data at an edge node's storage units.

The authors of [89] provide a polynomial-time greedy algorithm as a solution to the problem of data allocation in a group of heterogeneous mobile edge nodes. Their model takes into consideration the size of data, the storage capacity of the available nodes, a node-to-data demand matrix, as well as a corresponding transportation cost matrix. The result of the proposed model is an optimal decision for the data allocation and is computed at a (1-1/e)-approximation factor. This work presents an option that can easily be implemented at the edge of the network, owing to its efficiency and minimal demand of resources.

Moreover, the use of a system called Greedy Routing for Edge Data (GRED) is suggested by [104] for the efficient management of data storage and subsequent recovery. Data items and edge servers are mapped in a virtual space according to their IDs; afterwards, a data item is assigned to the closest edge server in the virtual space for storage. Data retrieval is implemented through the utilization of distributed hash tables (DHTs) for data stored at one-hop-away neighboring edge nodes in combination with SDN for query routing. Its experimental evaluation yields less than 30% routing cost and balances the load of data better than Chord [94], a popular DHT.

ECS (Edge-side Cooperative Storage) [40] constitutes a graph-based iterative algorithm that aims to place data that is required by an edge node to perform its tasks at its corresponding storage unit. The algorithm starts off by assigning to each edge server its most preferred data block and then continues by repetitively updating every node's assignment taking into consideration the other nodes' assignments. This framework can easily be implemented in a distributed manner and is not necessarily dependent on a centralized manager. One drawback is that an edge server is assigned with only one specific data block, whose size and content is predetermined, removing the possibility of an edge server storing a list of unrelated data items. Surely, that poses no problem unless an edge node performs many tasks at a time.

GAPSO [19] combines the best characteristics of the Genetic Algorithm (GA) and Particle Swarm Optimization (PSO) to distribute accordingly the data that are required from each edge node that participates in a scientific workflow. In particular, GA's crossover and mutation functions are integrated into PSO, whose ability to quickly converge to a solution is exploited. The decision is made considering the restrictions placed on the transmission costs. This algorithm involves little to no difficulty in implementation, while also achieving efficiency in performance.

Another approach to data management at the edge is represented in [41]; data can be stored at different edge nodes according to criteria like data solidity, which ensures that all data allocated to a specific node have a bounded standard deviation. This paper's authors utilize interpretable machine learning to compute the most important of an incoming data entry's features, and only based on them decide whether it will be stored locally or offloaded to a peer or the cloud. Each feature's importance is calculated with three different metrics and finalized by a neural network. Then, the most important features are imported to a Naïve Bayes model that determines the data entry's final location. Such a model can improve the time of retrieval for a range query, since data are stored in a quite sorted manner, hence proving itself valuable for an appropriate application setting.

### 5.2 Replica placement

A variety of models (like [5], [88]) are proposed for the prediction of data items' future demand and subsequent caching on edge nodes, should the corresponding demands be high. These frameworks allow for quick retrieval of requested data that are primarily stored on the Cloud. The authors of [58] suggest the use of two centralized algorithms for the allocation of data block replicas; DRC-GM is responsible for dynamically adjusting the number of replicas given the frequency of access to a specified data block, and RP-FNSG's goal is to find the optimal location for the aforementioned replicas in this problem of distributed placement. In [90], the authors introduce a model that combines the cloud and edge computing paradigms to place data replicas on both of the two infrastructures, thus delivering a time- and power-efficient solution for (Internet of Things) IoT scenarios that involve large datasets. More specifically, the problem is modeled as a 0-1 integer programming model and makes decisions through a variant of intelligent swarm optimization based on dependencies between data and the reliability of the storage hardware, among others. A content centric approach is the focus of [91] where a scheme is described around four complementary to each other algorithms for data caching at the edge. Each of these algorithms focuses on one of the following features to make latency minimum: data popularity, data heterogeneity, user mobility, and resource availability. Combining all these metrics to derive the final allocation makes for a powerful framework that can make effective decisions very fast. Indeed, the model was designed for use in a smart city, which requires efficient real-time decision-making above all else.

These approaches deliver positive results, seeing as data are in close proximity to a number of edge nodes that request them; however, replication techniques seem to be costlier than offloading models in terms of computational resources and latency. This is due to the involvement of the Cloud, which makes computations more complicated and communications slow.

### 5.3 Joint Data and Tasks Allocation

Data and tasks allocation/placement at the EC/EM is a significant research field/ A number of research activities try to detect the optimal location where data and tasks are distributed. The aim is to eliminate the time for delivering the final analytics while keeping maximized the 'matching' between requests and the available data. In [28], the authors focus on joint tasks and data placement over the edge of the network for data-intensive services. They suggest the use of a polynomial-time algorithm that solves their problem by treating it as a set function optimization. The solution takes into consideration resource constraints in computation, storage, communication, and cost. In [11], a scheme that allocates data to edge nodes and, then, if possible, allocates tasks to edge nodes that have in their storage the data they require. Specifically, the scheduler's data placement kernel can function in four ways: (i) no replication, where no data are replicated to an edge node before a request for them arrives, (ii) 1-replication, where each data block is replicated to only one edge node, (iii) full replication, where all data blocks are replicated and stored at each edge node in the network, and (iv) context-aware replication, where a strategy much like the one introduced in [58] is activated. Furthermore, the scheduler's task allocation component operates through (i) random task scheduling, (ii) data aware-scheduling, where a task is randomly placed on one of the edge nodes that possess a replica of the data required for the task's execution, and (iii) performance-aware scheduling, which places a task at the fastest edge node without a job that also possesses the required data at that moment. Such an approach is very appealing, but if the network it is applied to is large enough, transmission latencies risk exceeding the avoided cost of not having to communicate data very often. An additional research effort [55] proposes a scheme that seems superior to the previous two. First, data blocks are allocated to edge nodes based on (i) the value of the data, which depends on their popularity, the edge node's storage capacity and replacement ratio, denoting the data block's size in relation to the node's storage capacity, (ii) the transmission cost, and (iii) the replacement cost. The assignments are determined through the use of the Tabu search algorithm [31], [32]. After that comes the tasks allocation part, which is based on the allocated data blocks. A variant of the Kuhn-Munkres Hungarian method [99] is employed to find the optimal solution, given (i) the tasks' priorities, (ii) the relevance between tasks and data, and (iii) the transmission costs. This framework combines the two functionalities and, as experimental results show, achieves low response times, high hit rates and a low number replacements, making it an ideal solution for a generic joint data-and-task allocation scheme.

## 6. Tasks Management at the Edge

Since the number of owned personal devices has immensely increased these last few years, so have requests for services. The rise in the number of devices has made possible people's access to existing services and to new ones created to satiate the population's new needs. Service requests are now numerous and require much more processing power than before to be responded. EC/EM intends to bring the request processing closer to end users and reduce the bandwidth overload which restricts the performance of the Web, thus, lowering the related latency. However, executing tasks at the edge is not as simple as it sounds, owing to the heterogeneity of EC nodes and service requests at different locations. Placement/offloading techniques have to be employed to ensure the smooth operation of edge computing.

DCTA (Data-driven Cooperative Task Allocation) [16] was designed to manage multi-task transfer learning. It exploits the fact that each task has its own importance so as to produce an effective task allocation solution. In particular, the authors have observed that only a few of the requested tasks are actually important and, thus, DCTA assigns those tasks to the most powerful available edge nodes. To achieve that, at first, Clustered Reinforcement Learning (CRL), which is a combination of k-NN and Deep Q-learning, was created. However, due to simulation limitations that CRL poses, a cooperative learning approach that uses both CRL and SVM is finally proposed. Ultimately, even though the algorithm was developed to solve a quite specific problem, it can certainly be applied to the generic problem of tasks allocation, since the main idea of task importance applies to all tasks.

MobMig [78], a mobility-aware and migration-enabled approach, is a framework that consists of two algorithms for mobile edge computing tasks allocation. The first, mobility-aware allocation, is responsible of detecting incoming users in the range of an edge node that have unallocated task requests and then allocating them to edge nodes, based on a fitness function. The second, mobility-aware migration, aims to find overloaded edge nodes and relieve them by offloading tasks to nearby underloaded nodes. This scheme takes care of incoming tasks and redistributes existing tasks to reflect user mobility and solve node overloading, which can be a big problem if left untreated.

### 6.1 Placement of Containers

The authors of [61] focus on the migration of services to nearby edge nodes as the user is moving and approaching different nodes in a wide area network. They propose the use and migration of Docker containers, which they have studied extensively, and have been able to utilize their hierarchical file system to shorten the cost of its synchronization, all the while not depending on the distributed file system. The authors present experimental results that show an extensive decrease in service migration time, which renders this technique suitable for integration to any container allocation method. Further, the authors of [22] present two placement algorithms, KCBP and KCBP-WC, whose aim is to allocate container images to edge nodes while reducing the maximum transmission time of containers. To achieve that, it also exploits the hierarchical structure of containers and communicates only the layers that are not already in the receiving node's storage. The first algorithm, KCBP (k-Center-Based Placement) sorts layers by their size and places them at edge nodes so that the distance sum is minimized. KCBP-WC (KCBP-Without Conflict) is an extension that avoids placing two large layers at the same node, so that there is no extra overhead. Experimental results show a very large reduction in recovery time compared to other algorithms, supporting the proposed scheme's effectiveness.

## 6.2 Greedy Approaches

The authors of [2] have developed a greedy heuristic algorithm that assigns tasks to resources on the edge or the cloud for execution. Its criteria for the allocation include the deadline or other constraints connected to a task, the network distance between a task and a resource, and hence the latency it results in, a node's load of work and its capacity concerning energy. The algorithm's main objective is to produce a mapping of services to resources, while also minimizing the overall execution times and costs. The algorithm's time complexity equals $\theta(nm)$, where n is the number of tasks and m the number of nodes, which is a quite reasonable and expected value for a greedy algorithm. In [6], a number of algorithms for autonomous (re)placement of services in an edge node are described and compared to a centralized optimal MILP-based algorithm which minimizes the number of services that are delegated to the cloud for execution. In particular, the authors propose the admission and scheduling of service requests by an edge node only if the deadlines of already scheduled tasks as well as the new task are not violated; otherwise, the service request that the node denied is propagated to the cloud through the fog network, where it can also be admitted and executed by a fog node. The admitted requests are scheduled based on non-preemptive polices, such as Earliest Deadline First (EDF) and First-In-First-Out (FIFO). In addition, as a general rule, an edge cannot be running all services it receives requests. For that reason, it maintains a running set of services which is updated upon request arrival, periodically or at random intervals. This paper recommends the following ranking policies for the selection of running services: (i) Strictest Deadline First (SDF), where the requests of the services that are chosen have the least average amount of remaining time towards the deadline, (ii) Least Frequently Used (LFU), employed as an eviction policy for unpopular services, (iii) Hybrid, a mix of SDF and LFU, where services with distant-deadline requests are sent to the fog/cloud and the remaining are ranked based on popularity to the users, and (iv) Least Recently Used (LRU), which is also used as an eviction policy for services that have not been requested in a while. Out of the four, the Hybrid algorithm seems to be the best and closest in performance to the optimal. Such an approach presents an interesting option due to its autonomy, seeing as a node does not need to communicate with any other entity in order to make a decision. This results in an increased rate of decision-making and, generally, processing of requests, along with the improvement of the users' quality of service, which is the ultimate goal in any network.

A more specific approach [42] to tasks management with respect to a single edge node is to just decide on whether it is best for a particular task to be executed locally or offloaded to a peer node or the cloud. This paper's central idea involves building an incremental kernel density estimator for every task's requests as time progresses, that allows for the calculation of the probability of high request rates in the immediate future. Having exchanged this information among themselves, each edge node computes the probability of each task being popular generally in the group of nodes that does not include themself. Through a set of rules implemented by a fuzzy logic controller, an offloading value is calculated for each task by a node, and it indicates whether a task should remain for local execution or be offloaded elsewhere, where it is more popular to reduce the average latency for end users. This paper does not include the allocation of the task to a new edge node or the cloud, but the authors do refer to another work of theirs that does [46]. There, the Utility Theory is exploited to produce aggregated ranked lists of nodes where each task can be offloaded to, and if no node's related values exceed a certain threshold then the task is sent to the fog/cloud for execution that will not respect the task's deadline. Both of the proposed approaches' results are promising, as well as better compared to other distributed algorithms.

Overall, greedy approaches are dependent on the performance of each edge node. They work most efficiently if they require little to no interaction among them and certainly if interaction with the cloud is avoided. They can reduce the end user's waiting time since heavy computation is not employed to manage tasks. However, since an optimal solution is not usually possible, waiting time might sometimes reach very high values.

## 6.3 Energy-Saving Approaches

The authors of [59] study the placement problem from an IoT perspective and strive to develop an algorithm that finds the optimal solution to tasks management while also respecting edge servers' low energy reserve. They model the problem at hand as a multi-objective optimization problem and employ PSO to solve it and achieve the goal. Indeed, the algorithm does decrease the required energy, in comparison to other algorithms, but a centralized scheme such as the aforementioned can never accomplish optimal performance and energy consumption. In [86], it is proposed that the processing of heavy computations generated in an IoT environment be offloaded to other devices in their network once a specified level is reached. Instead of setting a universal offloading level for all devices, the authors suggest that each device has its own. In doing so, a network's devices' resources are utilized to a satisfactory extent and, at the same time, a bandwidth overload is avoided. Furthermore, experimental results show that this approach increases battery life, which is precious to IoT devices. This work is of the utmost importance to IoT networks which collect a large amount of data in need of aggregation or, in general, run applications requiring many computations.

DART (DAta tRansportation neTwork) [24] treats the tasks allocation problem as a stochastic mixed-integer optimization problem and attempts to solve it with an advanced coarse-grained offloading mechanism. The authors visualize IoT as a network of points of connection, which can communicate, compute tasks as well as store data and exploit the advantages of spectrum allocation in their model. Therefore, the devices in the network spend as little as possible of their much-needed energy for precise communication activities and save the rest for executing and offloading tasks. The authors of [60] strive to propose a model that utilizes data compression and task offloading to minimize a task's execution latency and the requesting device's energy consumption in a mobile edge computing environment. The problem is modeled as a non-convex one at first, but through conversion to a quadratically constrained quadratic program and then through a semidefinite relaxation approach, it is rendered solvable in polynomial time. Cooperatively optimizing task latencies and device energy exhaustion is a novel idea that seems quite suitable for devices with energy constraints.

## 6.4 Placement of Interacting Tasks

The Edge Orchestrator (EO) described in [79] is tasked with splitting a workflow into several parts to be executed across a number of devices and resources, as well as assigning tasks to be executed on an edge node or offloading the process for execution on a cloud. Furthermore, after a task has been performed at the edge, the EO must aggregate the resulting data before forwarding them to a cloud, and vice versa. As with most other models, the purpose is performance optimization with regard to latency and quality of service. The fact that this scheme offers a lot of alternatives for the allocation of requested tasks is positive since, out of all the possible solutions, the best one should be quite close to the optimal one. Having said that, the same feature must render the model slower to decide than others precisely because of the abundance of options.

Poster [37] aims to optimize communication costs between interacting tasks. To achieve that, it mines users' cookies with PISMine, an algorithm the authors have developed. PISMine finds the most common 2-itemsets of services based on an interestingness measure and places those that are a part of the same 2-itemset on the same edge node for execution, or as close as possible. Such a placement allows for the minimization of inter-service communication costs, thus minimizing the overall latency of a service and maximizing the users' quality of service (QoS). This approach is quite original and easy to implement, making it a very interesting choice for tasks management at the edge of the network.

The effort presented in [8] models the problem as a joint allocation of the multiple tasks an application might request. Similarly to [24], the problem is also modeled as a mixed integer linear problem (MILP) that takes into consideration user mobility and network capacity at a given moment. The authors present an effective heuristic online algorithm as a solution that is based on the Hungarian method [52]. Experiments show that the proposed approach is quick and makes virtually optimal decisions, making this scheme a very strong option for solving the tasks management problem at the edge.

DATA (Dependency-Aware Task Allocation) [53] is another approach at multi-component application allocation at the edge. This particular method is comprised of three sub-algorithms. The first one produces a graph that shows the order in which the tasks of a request must be processed, making sure that the all tasks in the graph are divided into sub-tasks which have the same maximum workload. This allows for the design of a pipeline of events, which is vital for the efficient operation of the next sub-algorithm. The second sub-algorithm's aim is to assign each sub-task to a container in the pool of available edge nodes in a manner that minimizes dependency-related transmission costs. Lastly, the third sub-algorithm is responsible of scheduling a sub-task's execution in the chosen container, by computing the maximum time that the inputs will have arrived at and respecting dependencies.

In addition, the authors of [103] suggest a scheme that utilizes RL, specifically Q-learning, to solve the problem of service request placement at the edge. A request is modeled as a set of sub-tasks, all of whom have to be allocated to edge nodes for processing. This paper introduces a strategy that allocates all the involved sub-tasks jointly, by treating them as a service tree with sub-trees which respect the order of events to take place. The model recursively allocates an edge node for every sub-tree using Q-learning, and eventually develops an efficient system that has learnt from its mistakes and can maximize the resource utility while minimizing the network congestion. This approach has favorable prospects as it can always evolve by adapting to changes in the network and maintain a high performance until a change occurs and it adapts to it again.

The services that are offered nowadays are complex and involve many components more often than not. Utilizing service placement methods which recognize that and strive for the minimization of inter-component latencies is an integral part of achieving overall efficiency in all networks, and especially a vast one like the world-wide web is.

Table 3 shows the optimization goals of each algorithm as well as the metrics that are taken into consideration. Symbol $\uparrow$ refers to the maximization of a goal, symbol $\downarrow$ to the minimization of a goal while symbol $\leftrightarrow$ depicts that algorithm accounts the specific metric for the final decisions. To avoid cluttering we merge the optimization goals and the accountable metrics for the first three columns. The first column (Resource Management) contains the following optimization parameters: (a) resource utilization, (b) recovery time, (c) execution cost, (d) load Balancing and (e) energy. The second column (Network Parameters) contains the following optimization parameters: (a) bandwidth, (b) topology, (c) network cost, (d) transmission time and (e) network traffic. Finally, column Data & Users Management contains: (a) data-users locality, (b) type of data, (c) data availability, (d) data solidity and (e) mobility. Consider the case of the algorithm presented in [33]. It can be easily revealed that the optimization goals of this algorithm are to minimize latency ($\downarrow$) and to maximize load balancing ($\uparrow^d$) and fault-tolerance ($\uparrow$). To do that it considers, the topology ($\leftrightarrow^b$) of the network, the locality, the type and the availability of the data ($\leftrightarrow^a \leftrightarrow^b \leftrightarrow^c$) and the storage capacity of the nodes ($\leftrightarrow$).

Table 4 depicts the classification of the algorithms for the following distinct perspectives: (a) the optimization strategy, (b) the appliance of replication methodologies, (c) the use of containers, (d) tasks dependencies (workflows), (e) the evaluation strategy and (f) the type of workloads used in the experimental evaluation. Consider again the case of the algorithm presented in [33]. This algorithm implements a heuristic approach to the SUMO simulator on real-world workloads.

**Table 3** Algorithms classification of optimization goals and accountable metrics

| Algorithm | Resource Management | Network Parameters | Data & Users Management | QoS | Execution Time | fault-tolerant | latency | storage | Cache-hits | Task importance |
|---|---|---|---|---|---|---|---|---|---|---|
| Gupta et al [33] | ↑[d] | ↔[b] | ↔[a] ↔[b] ↔[c] | | ↑ | | ↓ | ↔ | | |
| Ravindran et al [84] | | ↔[b] | | ↑ | | | ↓ | | | |
| Shao et al [89] | | ↔[c] | ↔[b] | | | | | | ↔ | ↑ |
| GRED [104] | ↑[d] | ↔[a] ↔[b] ↓[c] | | | | | ↓ | | | |
| ECS [40] | | | ↔[a] | | | | | ↔ | ↑ | |
| GAPSO [19] | | ↓[c] ↔[c] | | | | | ↓ | ↓ | | |
| Karanika et al [41] | | | ↔[b] ↑[d] ↔[d] | | | | ↓ | | | |
| Li et al [58] | ↔[d] | ↔[c] | ↔[c] | | ↔ | | ↓ | | | |
| Shao et al [90] | ↓[c] ↔[d] | ↔[c] | ↑[c] ↔[c] ↔[d] | | ↔ | | | ↑ | | |
| Sinky et al [91] | | ↓[e] | ↔[b] ↔[c] | ↑ | | ↔ | ↓ | | ↑ | |
| Farhadi et al [28] | ↔[c] | ↔[c] | | ↑ | ↓ | | | ↔ | | |
| Breitbach et al [11] | | ↓[a] ↔[e] | | | ↓ | | ↓ | | | |
| Li et al [55] | | ↓[a] ↔[c] | ↔[b] | ↑ | | | ↓ | ↔ | ↑ | |
| DCTA [16] | | | | | ↓ | | | | | ↔ |
| MobMig [78] | | ↔[b] | ↔[e] | | | | | | | |
| Ma et al [61] | | ↔[c] | ↑[e] | | | | ↓ | ↔ | ↔ | |
| KCBP [22] | ↓[b] | ↔[b] ↓[d] | | | | | | | | |
| Alqahtani et al [2] | ↓[c] ↔[e] | ↔[b] ↓[e] | | ↔ | | | ↓ | | | |
| Ascigil et al [6] | | ↔[b] | | ↑↔ | ↓ | | | | | |
| Karanika et al [42] | | ↔[b] | | | | | ↓ | | | ↔ |
| Kolomvatsos [46] | ↔[e] | ↔[b] | | | ↔ | | ↓ | | | ↔ |
| Li et al [59] | ↑[a] ↓[e] ↔[e] | ↔[b] | | | | | | | | |
| Samie, et al [86] | ↑[a] ↓[e] ↔[e] | | ↔[a] | | | | | | | |
| DART [24] | ↓[e] ↔[e] | ↔[b] | | | ↔ | | | | | |
| Ly et al [60] | ↓[e] ↔[e] | | ↓[d] ↔[d] | | ↔ | | ↓ | | | |
| Petri et al [79] | | | | ↔ | ↓ | | ↔ | | | |
| PISMine [37] | | ↓[c] | ↔[d] | ↑ | | | ↓ | | | |
| Bahreini et al [8] | | ↔[a] | ↔[a] | | | | | | | |
| DATA [53] | | ↔[d] | | | ↓↔ | | | | | |
| Wang et al [103] | ↑[a] | ↓[e] | | | ↔ | | | | | |

**Table 4** Algorithms classification for the scheduling and the application model

| Algorithm | Strategy | Replication | Containers | Dependency | Evaluation | | Workloads | |
|---|---|---|---|---|---|---|---|---|
| | | | | | Simulation | Real | Synthetic | Real |
| Gupta et al [33] | Heuristic | √ | | | √ SUMO | | | √ |
| Ravindran et al [84] | Heuristic | | √ | | √ NS3 | | √ | |
| Shao et al [89] | DP | | | | √ | | √ | |
| GRED [104] | Heuristic | | | | √ P4 | | √ | |
| ECS [40] | LP/IP | | | | √ BRITE | | | √ |
| GAPSO [19] | GA/PSO | | | √ | √CloudSim | | | √ |
| Karanika et al [41] | Ensemble ML | | | | √ | | | √ |
| Li et al [58] | GA | √ | | | | √Alibaba | | √ |
| Shao et al [90] | 0-1 IP/PSO | √ | √ | √ | | | √ | |
| Sinky et al [91] | Hierarchical Clustering | √ | | | | √ | | √ |

| Farhadi et al [28] | MILP | √ | | | √ | | √ | |
| Breitbach et al [11] | Heuristic | √ | √ | √Tasklet | | | | √ |
| Li et al [55] | Metaheuristic | √ ↓ | √ | | √ | | | √ |
| DCTA [16] | Heuristic, TL | | | | √ | | | √ |
| MobMig [78] | Heuristic, DM | | | | √ | | | √ |
| Ma et al [61] | Heuristic | √ | | | √ | | | √ |
| KCBP [22] | Heuristic | √ | √ | √ | | | √ | √ |
| Alqahtani et al [2] | Heuristic | | √ | √ViFogSim | √ | | | |
| Ascigil et al [6] | Heuristic | √ | | √ | | | √ | √ |
| Karanika et al [42] | Heuristic, FL | | | √ | | | | √ |
| Kolomvatsos at al [46] | k-NN | | | | | | √ | √ |
| Li et al [59] | PSO | | | √ | | | | |
| Samie, et al [86] | Heuristic | | | √DART | | √ | | |
| DART [24] | MILP | | | √ | | √ | | |
| Ly et al [60] | QCQP | | | √ | | √ | | |
| Petri et al [79] | Heuristic | | √ | | √CometCloud | | | √ |
| PISMine [37] | FP-tree | | | √ | | | | √ |
| Bahreini et al [8] | Heuristic | | | √ | | √ | | |
| DATA [53] | Heuristic | √ | | √ | | √ | | |
| Wang et al [103] | RL | √ | | √ | | √ | | |

## 7. Resources Management at the Edge

This section is devoted to the presentation of the most significant efforts that deal with the management of resources (e.g., processing nodes) at the EC/EM. Our aim is to reveal the axes that dictate the research implementations to deal with the requirements of the fully autonomous EC/EM infrastructure. Table 5 reports on the classification of the relevant research efforts according to the sub-domain.

Table 5. Categorization of research activities related to the management of EC/EM nodes.

| Research Subject | Short Description | Research Efforts |
|---|---|---|
| **Tasks Offloading** | | [17], [47], [98] |
| **Nodes reconfiguration** | | [4], [12], [21], [38], [43], [45], [50], [49] [57], [54], [63], [75], [93], [95], [106] |
| **Resources Scaling** | | [1], [23], [25], [26], [36], [39], [65], [66], [83], [87], [92], [96], [97], [101], [105], [107] |
| **Load balancing** | | [7], [18], [34], [56], [70], [71], [72], [73], [102] |

The subject of [17] is the intelligent computational offloading of mobile devices towards edge nodes, specifically at the edge of radio access networks. Task offloading to the available nodes is a significant research issue that demands for efficient solutions to increase the performance of the EC/EM. Tasks offloading should take into consideration nodes characteristics and the data present n them before the final decision. The activity can be coordinated by a multitude of EC/EM nodes accompanied by admission control and a specific scheduling scheme [98]. Additional efforts like [47] focus on the adoption of multiple criteria and try to 'match' tasks with EC/EM nodes characteristics before an allocation takes place. The discussed paper deals with the dynamic update of tasks and nodes characteristics due to the changes in the demand and the availability of nodes.

Nodes' reconfiguration is also very significant to keep the firmware and the pre-installed applications updated. Relevant efforts in the field [4], [43], [45], [49], [50] are mainly oriented to the IoT domain, however, the proposed algorithms can be easily adopted in EC/EM. A survey on the adopted methodologies is presented in [12]. Updates can affect the firmware of a device or have the form a 'generic' reprogramming activity. This means that an update server is adopted to deliver the updates that can be software patches, security modules, new functionalities/modules, etc. All the proposed algorithms try to minimize the effect in the network while targeting to limit the time for concluding the update. Incremental updates and data compression could be adopted to reduce the size of messages [95]. In any case, the separation of the updates impose additional requirements for the number of messages transferred through the network and the algorithm adopted to conclude the aggregation of messages and installation of the update. The incremental management of the updates does not eliminate the necessary process for maintaining updates history and the aforementioned aggregation. Some widely cited research efforts in the domain are as follows. Trickle [54] disseminates and, accordingly, maintains software updates in a set of nodes through an epidemic approach with scalable multicasting. Based on this scheme, updates are periodically transferred to nodes. Epidemic approaches, in general, may involve the transmission of several copies to random nodes, thus, there is an increased cost for the management of the received messages. DHV [21] reports on a code consistency maintenance protocol that ensures that nodes will, eventually, have the same code. The Multicast-based Code redistribution Protocol (MCP) [57] is another protocol that performs code maintenance. MCP requires a table that depicts the information of applications present in a node. The table is adopted for the 'coordination' of the delivery of multicast-based code dissemination requests. However, the use of additional data structures increase the storage complexity of the corresponding models. The Multi-hop, Over-the-Air code distribution Protocol (MOAP) [93] adopts a store-and-forward approach upon patterns of updates. Updates are broadcasted in a neighbour-per-neighbour basis forcing nodes to disseminate the incoming code to reduce the latency. Deluge [38] proposes a protocol over algorithms related to density-aware, epidemic maintenance models. It is built upon Trickle for the advertisement of updates and separates the code into a set of fixed-size pages. Through this approach, the time required for the propagation of large components is reduced. However, the adoption of multiple optimization activities increases the complexity of the proposed solution especially for the recreation of the updates. Stream [75]adopts Deluge and optimizes the code parts transferred through the network. It deals with pre-installations of the re-programming application. This way, Stream transmits the minimal support (approximately one page) required for the activation of the re-programming image. Resource-awareness, time-efficiency, and the integration of security solutions are involved in the model presented in [63]. A multi-hop propagation scheme is proposed enhanced by security codes and means from fuzzy control theory. MELETE [106] is designed to support multiple concurrent applications. It assumes that the network is a set of groups of nodes that execute different tasks. The framework adopts a group-keyed model to selectively distribute the code to only the interested nodes, and reactively distribute the code only when it is required.

The dynamic scaling activities for EC/EM nodes is another significant resources topic. Scaling activities can be performed either horizontally [1] or vertically [92]. Horizontal scaling refers in adding/removing infrastructure capacity in pre-packaged blocks of resources while vertical scaling refers in scale-up/scale-down, i.e., add/remove resources to an existing system [92], [101]. Scaling activities can deal with the management of Virtual Machines (VMs) [105] or containers [87], [107]. Activities adopted to aligned with real demand can also involve the migration of services or data to different VMs/containers [65]. ENORM [101] proposed a framework for integrating the EC/EM in the computing ecosystem to realize FC. The framework builds on a provisioning and deployment model to integrate an EC/EM node with a Cloud server. Additionally, it supports an auto-scaling mechanism to dynamically manage edge resources to be fully aligned with the real demand. DYVERSE [100] is a light-weight and dynamic vertical scaling mechanism for managing resources allocated to applications for facilitating multi-tenancy in EC/EM. DYVERSE proposes the use of a static and three dynamic priority management scheme being workload-, community- and system-aware. Thoth [87] proposes a dynamic resource management system using Docker container technology. It automatically monitors resource usage and dynamically adjusts appropriate amount of resources for each application based on ML models, i.e., a Neural Network, a Q-Learning scheme and a rule-based algorithm. Other efforts that adopt reinforcement learning for predicting the demand and align the available resources are discussed in [25], [83]. A fuzzy logic approach is proposed in [39]. The fuzzy controller is adopted to result the scaling actions based on demand prediction estimated by a reinforcement learning scheme. In [26], the authors present a queuing mathematical and analytical model to study and analyze the performance of fog computing system. The discussed model determines under any workload the number of nodes required to keep the QoS at the desired levels. In [97], the proposed scheme integrates hypervisors and virtualization based on containers to construct an integrated virtualization platform for industrial applications. The adopted model is a fuzzy-based real-time autoscaling mechanism that provides a dynamic, rapid, lightweight, and low-cost solution.

The authors of [36] study the technical challenges for managing the resource-limited nodes in EC/EM. They present three architectures, i.e., dataflow, control, and tenancy. The infrastructure is seen in three axes, i.e., hardware, software, and middleware. They also discuss algorithms for load balancing, discovery, benchmarking, and placement. Another effort studying the resources continuity is provided in [66]. The study focuses on the management of resource continuity from the EC/EM to Cloud and depicts a layered architecture for continuity provisioning, effective resources selection and service execution mechanisms. We have to notice that resources should be able to be aligned with the needs of streams as coming from the IoT infrastructure. The authors of [23] provide a survey on stream processing techniques for supporting resource elasticity features. The review focuses on ongoing efforts dealing with the deployment on EC/EM environments and insights of future directions. A taxonomy of resource management at the EC/EM is provided by [96]. The authors categorize the relevant efforts according to their resource type, the objective of the management, resource location, and resource use.

Load balancing is another significant research subject in EC/EM. The research community proposes two main types of load balancing strategies: static and dynamic [56]. Static load balancing deals with a 'stateless' approach (models that do not consider the previous state of the node). The load is efficiently distributed when nodes do not exhibit significant variations in their activities. Dynamic strategies involve a 'statefull' model, i.e., they take into consideration the dynamic changes in nodes' behavior (the state of each node). Intermediary nodes can be also adopted for performing load balancing activities [7]. In [56], the authors propose an architecture for load balancing based on intermediary nodes that obtain the state information of the network. Intermediaries classify the status of each node by using a set of attributes. Based on this information, the framework is capable of performing the final allocation. In [34], the authors propose an improved constrained particle swarm optimization algorithm based on SDN. The algorithm improves the performance by adopting the opposite property of the mutated particles and reducing the inertia weight linearly. In [18], a task allocation model is provided for load balancing. The algorithm calculates the completion time for each task and formulates the load balancing optimization problem. The authors of [102] present a distributed traffic management system adopting an offloading algorithm for real-time traffic management in fog-based internet of vehicle systems. The ultimate goal is to minimize the average response time of the traffic management server. In [73], the authors investigate a joint computation offloading, power allocation, and channel assignment scheme for 5G-enabled traffic management systems. The satisfaction of heterogeneous requirements for communications,, computation and storage are studied in [72]. The authors propose an energy-efficient scheduling framework taking into consideration task latency constraints. In [71], a deep learning model is presented for data transmission. The use cases involve the communication of vehicles with the EC/EM infrastructure. In [70] another deep leaning model is presented. More specifically, the authors focus on a deep reinforcement learning method integrated with vehicular edge computing.

8. Conclusions

The Edge Computing (EC) and the Edge Mesh (EM) target to support an intelligent infrastructure close to end users realizing services that can be applied upon the huge volumes of data collected by numerous devices. The ultimate (and first) goal is the minimization of the latency that users/applications enjoy when requesting the execution of various processing activities. The challenges that should be met before we arrive at a fully automated EC/EM infrastructure are many. The research community has already started to provide solutions to a wide set of problems relevant to this vision. Our paper serves the goal of presenting such efforts working mainly around the data, tasks and resources management. We want to reveal the paths for supporting additional services and models starting from a concrete basis. We discuss and classify a high number of models while categorizing them upon their research target. We present how legacy and advanced ML as well as optimization techniques are adopted to support novel solutions while reporting on the pros and cons of the discussed algorithms. We argue that all these aspects are parts of the same picture, i.e., to conclude an intelligent edge node that is capable of learning the status of itself, its peers and the environment before taking any decision about the upcoming line of actions. This means that we want to reveal the need of an adaptive node that reasons upon the contextual information of everything. If this becomes true, we will able to see a fully automated EC/EM ecosystem that is not a science fiction but a reality with huge benefits for services provided to end users.

**References**

[1] Ali-Eldin, A., Tordsson, J., Elmroth, E., 'An adaptive hybrid elasticity controller for cloud infrastructures', in IEEE Network Operations and Management Symposium, 2012, pp. 204–212.

[2] Alqahtani, D. N. Jha, P. Patel, E. Solaiman and R. Ranjan, "SLA-aware Approach for IoT Workflow Activities Placement based on Collaboration between Cloud and Edge," in 1st Workshop on Cyber-Physical Social Systems (CPSS) 2019, Newcastle, UK, 2019.

[3] Al-Qamash, A., Soliman, I., Abulibdesh, R., Moutaz, S., 'Cloud, Fog, and Edge Computing: A Software Engineering Perspective', International Conference on Computer and Applications (ICCA), 2018.

[4] Anagnostopoulos, C., Kolomvatsos, K., 'An Intelligent, Time-Optimized Monitoring Scheme for Edge Nodes', Journal of Network and Computer Applications, Elsevier, vol. 148, 2019.

[5] Aral and T. Ovatman, "A decentralized replica placement algorithm for edge computing," IEEE transactions on network and service management, vol. 15, no. 2, pp. 516-529, 2018.

[6] Ascigil, O., et al., "On Uncoordinated Service Placement in Edge-Clouds," in 2017 IEEE International Conference on Cloud Computing Technology and Science (CloudCom), Hong Kong, China, 2017.

[7] Babu, R., Joy, A., Samuel, P., 'Load balancing of tasks in cloud computing environment based on bee colony algorithm', 5th International Conference on Advances in Computing and Communications (ICACC), 2015, pp. 89–93.

[8] Bahreini, T., Grosu, D., "Efficient placement of multi-component applications in edge computing systems," in SEC '17: Proceedings of the Second ACM/IEEE Symposium on Edge Computing, San Jose, CA, USA, 2017, October.

[9] Basir, R., et al., 'Fog Computing Enabling Industrial Internet of Things: State-of-the-Art and Research Challenges', Sensors, MDPI, 19, 2019.

[10] Bellman, R., "Dynamic programming and stochastic control processes," Information and control, vol. 1, no. 3, pp. 228-239, 1958.

[11] Breitbach, M. et al., "Context-Aware Data and Task Placement in Edge Computing Environments," in 2019 IEEE International Conference on Pervasive Computing and Communications (PerCom), Kyoto, Japan, 2019.

[12] Brown, S., Sreenan, C., 'Software Updating in Wireless Sensor Networks: A Survey and Lacunae', Journal of Sensor and Actuators, vol. 2, 2013, pp. 717-760.

[13] Buhmann, J., Kuhnel, H., "Unsupervised and supervised data clustering with competitive neural networks," in IJCNN International Joint Conference on Neural Networks, 1992.

[14] Cao, H. et al., 'Analytics Everywhere: Generating Insights From the Internet of Things', IEEE Access, vol. 7, 2019, pp. 71749 – 71769.

[15] Carpa, M., et al., 'Edge Computing: A Survey on the Hardware Requirements in the Internet of Things World', Future Internet, MDPI, 11, 2019.

[16] Chen, Q., et al., "Data-driven task allocation for multi-task transfer learning on the edge," in 2019 IEEE 39th International Conference on Distributed Computing Systems (ICDCS), Dallas, TX, USA, 2019.

[17] Chen, X., Jiao, L., Li W., Fu, X., 'Efficient Multi-User Computation Offloading for Mobile-Edge Cloud Computing', IEEE/ACM Transactions on Networks, 2016, 24, 2795–2808.

[18] Chen, Y. A., Walters, J., Crago, S., 'Load balancing for minimizing deadline misses and totalruntime for connected car systems in fog computing', IEEE International Symposium on Parallel and Distributed Processing with Applications, 2017.

[19] Chen, Z., et al., "Effective data placement for scientific workflows in mobile edge computing using genetic particle swarm optimization," Concurrency and Computation: Practice and Experience, p. e5413, 2019.

[20] Cisco, 'The Cisco Edge Analytics Fabric System', White paper, 2016

[21] Dang, T., Bulusu, N., Feng, W., Park, S., 'DHV: A Code Consistency Maintenance Protocol for Wireless Sensor Networks', In Proceedings of the 6th European Conference on Wireless Sensor Networks, Cork, Ireland, 2009.

[22] Darrous, J., Lambert, T., Ibrahim, S., "On the Importance of Container Image Placement for Service Provisioning in the Edge," in 2019 28th International Conference on Computer Communication and Networks (ICCCN), Valencia, Spain, 2019.

[23] Dias de Assunção, M., da Silva Veith, A., Buyya, R., 'Distributed data stream processing and edge computing: A survey on resource elasticity and future directions;, Journal of Networks Computing Applications, 103, 1–17, 2018.

[24] Ding, H., et al., "Beef Up the Edge: Spectrum-Aware Placement of Edge Computing Services for the Internet of Things," IEEE Transactions on Mobile Computing, vol. 18, no. 12, pp. 2783-2795, 2019.

[25] Dutreilh, X., Kirgizov, S., Melekhova O., Malenfant, J., Rivierre, N. and Truck, I., 'Using Reinforcement Learning for Autonomic Resource Allocation in Clouds: Toward a Fully Automated Workflow', 7th International Conference on Autonomic and Autonomous Systems, 2011, pp.67-74.

[26] El Kafhali, S., Salah, K., 'Efficient and dynamic scaling of fog nodes for IoT devices', Journal of Supercomputing, 73(12), 2017, 5261–5284.

[27] ETSI, Mobile-edge Computing Introductory Technical White Paper, White Paper, Mobile-edge Computing Industry Initiative, 2014.

[28] Farhadi, V., et al.,, "Service Placement and Request Scheduling for Data-intensive Applications in Edge Clouds," in IEEE INFOCOM 2019-IEEE Conference on Computer Communications, Paris, France, 2019.

[29] Farzad, S., Bauer, L., Henkel, J., 'New Problems and Challenges in Bandwidth Allocation for IoT', Internet of Things Symposium, Amsterdam, Netherlands, 2015.

[30] Foukas, X., Patounas, G., Elmokashfi, A., Marina, M., 'Network slicing in 5G: Survey and challenges', IEEE Communications Magazine, 55(5): 94-100, May 2017.

[31] Glover, F., "Tabu Search—Part I," ORSA Journal on Computing, vol. 1, no. 3, pp. 190-206, 1989.

[32] Glover, F., "Tabu search—part II," ORSA Journal on computing, vol. 2, no. 1, pp. 4-32, 1990.

[33] Gupta, H., Xu, Z., Ramachandran, U., "DataFog: Towards a Holistic Data Management Platform for the IoT Age at the Network Edge," in {USENIX} Workshop on Hot Topics in Edge Computing (HotEdge 18), Boston, MA, USA, 2018.

[34] He, X., Ren, Z,. Shi, C., Jian, F., 'A novel load balancing strategy of software-defined cloud/fog networking in the internet of vehicles', Chinese Communications, 13(S2), 145–154, 2016.
[35] Hebb, D. O.,The organization of behavior: a neuropsychological theory, New York, NY: Wiley, 1949.
[36] Hong, C.-H. and B. Varghese, 'Resource Management in Fog/Edge Computing: A Survey', arXiv preprint arXiv:1810.00305, 2018.
[37] Huang, Y., et al., "Poster: Interacting Data-Intensive Services Mining and Placement in Mobile Edge Clouds," in Proceedings of the 23rd Annual International Conference on Mobile Computing and Networking, London, United Kingdom, 2017.
[38] Hui, J. W., Culler, D., 'The dynamic behavior of a data dissemination protocol for network programming at scale', in Procedings of the International Conference on Embedded networked sensor systems, SenSys, 2004, pp. 81-94.
[39] Jamshidi, P., Sharifloo, A., Pahl, C., Arabnejad, H., Metzger, A. Estrada, G., 'Fuzzy Self-Learning Controllers for Elasticity Management in Dynamic Cloud Architectures', 12th International ACM SIGSOFT Conference on Quality of Software Architectures, 2016, pp. 70-79.
[40] Jin, J., Li, Y., Luo, J., "Cooperative storage by exploiting graph-based data placement algorithm for edge computing environment," Concurrency and Computation: Practice and Experience, vol. 30, no. 20, p. e4914, 2018.
[41] Karanika, P. Oikonomou, K. Kolomvatsos and C. Anagnostopoulos, "An Ensemble Interpretable Machine Learning Scheme for Securing Data Quality at the Edge," in Cross Domain Conference for Machine Learning and Knowledge Extraction (CD-MAKE 2020), 2020.
[42] Karanika, P. Oikonomou, K. Kolomvatsos and T. Loukopoulos, "A Demand-driven, Proactive Tasks Management Model at the Edge," in IEEE International Conference on Fuzzy Systems, 2020.
[43] Kolomvatsos, K., 'An Efficient Scheme for Applying Software Updates in Pervasive Computing Applications', Journal of Parallel and Distributed Computing, Elsevier, vol. 128, 2019, pp. 1-14.
[44] Kolomvatsos, K., 'An Intelligent, Uncertainty Driven Management Scheme for Software Updates in Pervasive IoT Applications', Elsevier Future Generation Computer Systems, vol. 83, pp. 116-131, 2018.
[45] Kolomvatsos, K., 'An Intelligent, Uncertainty Driven Management Scheme for Software Updates in Pervasive IoT Applications', Elsevier Future Generation Computer Systems, vol. 83, pp. 116-131, 2018.
[46] Kolomvatsos, K., Anagnostopoulos, C., "Multi-criteria optimal task allocation at the edge," Future Generation Computer Systems, vol. 93, pp. 358-372, 2019.
[47] Kolomvatsos, K., Anagnostopoulos, C., 'Multi-criteria Optimal Task Allocation at the Edge', Elsevier Future Generation Computer Systems, vol. 93, 2019, pp. 358-372.
[48] Kolomvatsos, K., 'Time-Optimized Management of IoT Nodes', Elsevier Ad Hoc Networks, vol. 69, 2018, pp. 1-14.
[49] Kolomvatsos, K., 'Time-Optimized Management of IoT Nodes', Elsevier Ad Hoc Networks, vol. 69, 2018, pp. 1-14.
[50] Kolomvatsos, K., 'Time-Optimized Management of Mobile IoT Nodes for Pervasive Applications', Journal of Network and Computer Applications, Elsevier, vol. 125, 2019, pp. 155-167.
[51] Kolomvatsos, K.,, 'An Efficient Scheme for Applying Software Updates in Pervasive Computing Applications', Journal of Parallel and Distributed Computing, Elsevier, vol. 128, 2019, pp. 1-14.
[52] Kuhn, H. W., "The Hungarian method for the assignment problem," Naval research logistics quarterly, vol. 2, no. 1-2, pp. 83-97, 1955.
[53] Lee, J., et al., "DATA: Dependency-Aware Task Allocation Scheme in Distributed Edge Cloud," IEEE Transactions on Industrial Informatics, 2020.
[54] Levis, P, Patel, N., Culler, D., Shenker, S., 'Trickle: a self-regulating algorithm for code propagation and maintenance in wireless sensor networks', in Proceedings of the Symposium on Networked Systems Design and Implementation, vol. 1, 2004.

[55] Li, C., Bai, J., Tang, J., "Joint optimization of data placement and scheduling for improving user experience in edge computing," Journal of Parallel and Distributed Computing, vol. 125, pp. 93-105, 2019, March.
[56] Li, G., Yao, Y., Wu, J., Liu, X., Sheng, X., Lin, Q., 'A new load balancing strategy by task allocation in edge computing based on intermediary nodes', EURASIP Journal on Wireless Communications and Networking volume 2020.
[57] Li, W., Zhang, Y., Childers, B., 'MCP: an Energy-Efficient Code Distribution Protocol for Multi-Application WSNs', in Proceedings of the 5th IEEE International Conference on Distributed Computing in Sensor Systems, 2009.
[58] Li, Y. Wang, H. Tang, Y. Zhang, Y. Xin and Y. Luo, "Flexible replica placement for enhancing the availability in edge computing environment," Computer Communications, vol. 146, pp. 1-14, 2019.
[59] Li, Y., Wang, S., "An energy-aware edge server placement algorithm in mobile edge computing," in 2018 IEEE International Conference on Edge Computing (EDGE), San Francisco, CA, USA, 2018, July.
[60] Ly, M. H., Dinh, T. Q., Kha, H. H., "Joint Optimization of Execution Latency and Energy Consumption for Mobile Edge Computing with Data Compression and Task Allocation," in 2019 International Symposium on Electrical and Electronics Engineering (ISEE), Ho Chi Minh, Vietnam, 2019.
[61] Ma, L., Yi, S., Li, Q., "Efficient Service Handoff Across Edge Servers via Docker Container Migration," in SEC '17: Proceedings of the Second ACM/IEEE Symposium on Edge Computing, San Jose, CA, USA, 2017, October.
[62] Mahdavinejad, M. S., Rezvan, M., Barekatain, M., Adibi, P., Barnaghi, P., Sheth, A., 'Machine learning for internet of things data analysis: a survey', Digital Communications and Networks, 4(3):161 – 175, 2018.
[63] Maier, K., Hessler, A., Ugus, O., Keller, J., Westhoff, D., 'Multi-Hop Over-The-Air Reprogramming of Wireless Sensor Networks using Fuzzy Control and Fountain Codes', in Self-Organising, Wireless Sensor and Communication Networks, 2009.
[64] Manyika, J., Chui, M., Bisson, P., Woetzel, J., Dobbs, R., Bughin, J., Aharon, D., 'The Internet of Things: Mapping the Value Behind the Hype', Technical report, McKinsey Global Institute, 2015.
[65] Mao, M., Humphrey, M., 'Auto-scaling to minimize cost and meet application deadlines in cloud workflows', International Conference on High Performance Computation, Networking, Storage and Analysis, 2011, pp. 1–12.
[66] Masip-Bruin, X., Marin-Tordera, E., Jukan, A., Ren, G.J., 'Managing resources continuity from the edge to the cloud: architecture and performance', Future Generation Computer Systems, 79, 777–785, 2018.
[67] Miraz, H. M., Ali, M., Picking, R., 'A review on Internet of Things (IoT), Internet of Everything (IoE) and Internet of Nano Things (IoNT)', Internet Technologies and Applications (ITA), 2015.
[68] Mordor Intelligence, 'Micro Server Market – Growth, Trends, and Forecast (2020-2025)', Report, January 2020.
[69] Nebbiolo Technologies Inc., 'Fog vs Edge Computing', White paper, retrieved July 2020 by https://www.nebbiolo.tech/wp-content/uploads/whitepaper-fog-vs-edge.pdf
[70] Ning, Z, Dong, P., Wang, X., Rodrigues, J., Xia, F., 'Deep reinforcement learning for vehicular edge computing: an intelligent offloading system', ACM Transactions on Intelligent Systems Technology, 2019.
[71] Ning, Z., Feng, Y., Kong, X., Guo, L., Hu, X., Bin, H., 'Deep learning in edge of vehicles: exploring trirelationship for data transmission', IEEE Transactions on Industrial Informatics, 2019.
[72] Ning, Z., Huang, J., Wang, X., Rodrigues, J., Guo, L., 'Mobile edge computing-enabled internet of vehicles: toward energy-efficient scheduling', IEEE Networks, 2019

[73] Ning, Z., Wang, X., Xia, F., Rodrigues, J., 'Joint computation offloading, power allocation, and channel assignment for 5g-enabled traffic management systems', IEEE Transactions on Industrial Informatics, 2019.
[74] Odun-Ayo, I., Okereke, C., Orovwode, H., 'Cloud Computing and Internet of Things: Issues and Developments', in World Congress on Engineering, London, UK, 2018.
[75] Panta, R., Khalil, I., Bagchi, S., 'Stream: Low overhead wireless reprogramming for sensor networks', in Proceedings of the International Conference on Computer Communications, INFOCOM, 2007, pp. 928-936.
[76] Parker, M., 'Implementation with GPUs', Digital Signal Processing, 2017, pp. 387-393
[77] Patel, K. K., Patel, S. M., 'Internet of Things-IoT: Definition, Characteristics, Architecture, Enabling Technologies, Application & Future Challenges', International Journal of Engineering Science and Computing, vol. 6(5), 2016.
[78] Peng, Q., et al., "Mobility-Aware and Migration-Enabled Online Edge User Allocation in Mobile Edge Computing," in 2019 IEEE International Conference on Web Services (ICWS), Milan, Italy, 2019.
[79] Petri, O. Rana, A. R. Zamani and Y. Rezgui, "Edge-Cloud Orchestration: Strategies for Service Placement and Enactment," in 2019 IEEE International Conference on Cloud Engineering (IC2E), Prague, Czech Republic, 2019.
[80] Popescu, D., Zilberman, N., Moore, A. W., 'Characterizing the Impact of Network Latency on Cloud-based Applications Performance', Computer Laboratory technical reports, UCAM-CL-TR-914, 2017.
[81] Psaras, O. Ascigil, S. Rene, G. Pavlou, A. Afanasyev and L. Zhang, "Mobile Data Repositories at the Edge," in {USENIX} Workshop on Hot Topics in Edge Computing (HotEdge 18), 2018.
[82] Rahman, H., Rahmani, R., 'Enabling Distributed Intelligence Assisted Future Internet of Things Controller (FITC)', Applied Computing and Informatics, 14(1), 2018, pp. 73-87.
[83] Rao, J., Bu, X., Xu, C.Z., Wang, L., and Yin, G., 'VCONF: a reinforcement learning approach to virtual machine auto-configuration', 6th International Conference on Autonomic Computing, 2009, pp. 137-146
[84] Ravindran and A. George, "An Edge Datastore Architecture For Latency-Critical Distributed Machine Vision Applications," in {USENIX} Workshop on Hot Topics in Edge Computing (HotEdge 18), Boston, MA, USA, 2018.
[85] Sahni, Y., et al., 'Edge Mesh: A New Paradigm to Enable Distributed Intelligence in Internet of Things', IEEE Access, 2017
[86] Samie, F., et al., "Computation offloading and resource allocation for low-power IoT edge devices," in 2016 IEEE 3rd World Forum on Internet of Things (WF-IoT), 2016.
[87] Sangpetch, A., Sangpetch, O., Juangmarisakul, N., Warodom, S., 'Thoth: Automatic resource management with machine learning for container-based cloud platform', International Conference on Cloud Computing and Services Science, 2017, pp. 103–111.
[88] Saputra, Y. M. et al., "Distributed deep learning at the edge: A novel proactive and cooperative caching framework for mobile edge networks," IEEE Wireless Communications Letters, vol. 8, no. 4, pp. 1220-1223, 2019.
[89] Shao, X., et al., "A Competitive Approximation Algorithm for Data Allocation Problem in Heterogenous Mobile Edge Computing," in 2019 IEEE 89th Vehicular Technology Conference (VTC2019-Spring), 2019.
[90] Shao, Y., Li, C., Tang, H., "A data replica placement strategy for IoT workflows in collaborative edge and cloud environments," Computer Networks, vol. 148, p. 46–59, 2019.
[91] Sinky, H., et al., "Adaptive Edge-Centric Cloud Content Placement for Responsive Smart Cities," IEEE Network, vol. 33, no. 3, pp. 177-183, 2019.
[92] Sotiriadis S., Bessis N., Amza C., Buyya R., 'Vertical and horizontal elasticity for dynamic virtual machine reconfiguration', IEEE Transactions on Service Computing, vol. 99, 2016.

[93] Stathopoulos, T., Heidemann, J., Estrin, D., 'A remote code update mechanism for wireless sensor networks', Technical Report, Center for Embedded Networked Sensing, 2003.

[94] Stoica, R. Morris, D. Karger, M. F. Kaashoek and H. Balakrishnan, "Chord: A scalable peer-to-peer lookup service for internet applications," ACM SIGCOMM Computer Communication Review, vol. 31, no. 4, pp. 149-160, 2001.

[95] Stolikj, M., Cuijpers, P.,. Lukkien, J., 'Efficient Reprogramming of Wireless Sensor Networks Using Incremental Updates and Data Compression', in Proceedings of the IEEE International Conference on Pervasive Computing and Communications Workshops, 2013, pp. 584-589.

[96] Toczé, K., Nadjm-Tehrani, S., 'A taxonomy for manage-ment and optimization of multiple resources in edge computing', Wireless Communications and Mobile Computing, 2018, 1–23.

[97] Tseng, F.-H., Tsai, M.S., Tseng, C.W., Yang, Y.T., Liu, C.C., Chou, L.D., 'A lightweight auto-scaling mechanism for fog computing in industrial applications', IEEE Transactions on Industrial Informatics, 14(10), 2018, 4529–4537.

[98] Wang, L., Jiao, L., Kliazovich, D., Bouvry, P., 'Reconciling task assignment and scheduling in mobile edge clouds', In Proceedings of the IEEE 24th International Conference on Network Protocols (ICNP), Singapore, 2016.

[99] Wang, N., Fei, Z., Kuang, J., "QoE-aware Resource Allocation for Mixed Traffics in Heterogeneous," in IEEE International Conference on Communication, Shenzhen, China, 2017, 14-16 December.

[100] Wang, N., Matthaiou, M., Nikolopoulos, D., Varghese, B., 'DYVERSE: DYnamic VERtical Scaling in multi-tenant Edge Environment', Future Generation Computer Systems, vol. 108, 2020, pp. 598-612.

[101] Wang, N., Varghese, B., Matthaiou, M., & Nikolopoulos, D., 'ENORM: A Framework For Edge NOdeResource Management', IEEE Transactions on Services Computing, 2017.

[102] Wang, X., Ning, Z., Wang, L., 'Offloading in internet of vehicles: A fog-enabled real-time traffic management system', IEEE Transactions on Industrial Informatics, 2018.

[103] Wang, Y., et al., "A Reinforcement Learning Approach for Online Service Tree Placement in Edge Computing," in 2019 IEEE 27th International Conference on Network Protocols (ICNP), Chicago, IL, USA, 2019, October.

[104] Xie, J., et al., , "Efficient Data Placement and Retrieval Services in Edge Computing," in 2019 IEEE 39th International Conference on Distributed Computing Systems (ICDCS), 2019.

[105] Yazdanov, L., Fetzer, C., 'Lightweight automatic resource scaling for multi-tier web applications', in IEEE International Conference on Cloud Computing, 2014, pp. 466–473.

[106] Yu, Y., Rittle, L. J., Bhandari, V., Lebrun, J. B., 'Supporting concurrent applications in wireless sensor networks', in Proceedings of the 4th International Conference on Embedded Networked Sensor systems, SenSys, 2006.

[107] Zhang F., Tang X., Li X., Khan S.U., Li Z., 'Quantifying cloud elasticity with container-based autoscaling', Future Generation Computer Systems, 98, 2019, pp. 672-681.